\documentclass[twocolumn]{aa}
\usepackage{graphicx}
\usepackage{natbib}

\usepackage{txfonts}
\begin{document}
   \title{ The first high-resolution X-ray spectrum of a Herbig Star: The case of AB Aurigae}


   \author{Alessandra Telleschi
          \inst{1},
          Manuel G\"udel \inst{1},
          Kevin R. Briggs \inst{1},
          Stephen L. Skinner \inst{2},
          Marc Audard \inst{3},
          \and
          Elena Franciosini \inst{4}
          }

   \authorrunning{A. Telleschi et al.}
   \offprints{A. Telleschi}

   \institute{Paul Scherrer Institut, W\"urenlingen and Villigen,
              CH-5232 Villigen PSI, Switzerland\\
              \email{atellesc@astro.phys.ethz.ch}
         \and
             Center for Astrophysics and Space Astronomy, University of Colorado, Boulder,
             CO 80309-0389 
             \and
             Columbia Astrophysics Laboratory, Mail Code 5247, 550 West 120th Street, New York, NY 10027
             \fnmsep\thanks{\emph{New address (since September 2006): }Integral Science Data Centre, Ch. d'Ecogia 16, CH-1290 Versoix, Switzerland \& Geneva Observatory, University of Geneva, Ch. des Maillettes 5
1, 1290 Sauverny, Switzerland}
             \and
             INAF - Osservatorio Astronomico di Palermo, Piazza del Parlamento 1, 90134 Palermo, Italy}

   \date{Received 13 April 2006; accepted 13 October 2006}

  \abstract
   {The X-ray emission from Herbig Be/Ae stars has so far defied an unequivocal explanation. In later-type T Tauri
   stars, X-rays are thought to be produced by magnetically trapped coronal plasma, although accretion-shock
   induced X-rays have also been suggested. { In earlier-type (OB) stars, shocks in unstable winds are thought
   to produce X-rays.}}
   {We present the first high-resolution X-ray spectrum of a prototypical Herbig star (AB Aurigae),
   measure and interpret various spectral features, and compare our results with model predictions.}
   {We use X-ray spectroscopy data from the  XMM-Newton Reflection Grating Spectrometers and the EPIC
   instruments. The spectra are interpreted using thermal, optically thin emission models with variable
   element abundances and a photoelectric absorption component. We interpret line flux ratios in He-like triplet
   of O\,{\sc vii} as a function of electron density and the UV radiation field. We use the
   nearby co-eval classical T Tauri star SU Aur as a comparison. }
   {AB Aurigae reveals a  soft X-ray spectrum, most plasma being concentrated 
   at 1--6~MK. The He-like triplet reveal no signatures of increased densities 
   as reported for some accreting T Tau stars in the previous literature. 
   There are also no clear indications for strong abundance anomalies in 
   the emitting plasma. The light curve displays modulated variability,
   with a period of $\approx$ 42 hr. }
  {It is unlikely that a nearby, undetected lower-mass companion is the source of 
  the X-rays. Accretion shocks close to the star should be irradiated by the 
  photosphere, leading to alteration in the He-like triplet fluxes of  O\,{\sc vii},
  which we do not measure. Also, no indications for high densities are found, 
  although the mass accretion rate is presently unknown. Emission from wind 
  shocks is unlikely, given the weak radiation pressure. A possible explanation 
  would be a solar-like magnetic corona. Magnetically confined winds provide 
  a very promising alternative. The X-ray period is indeed close to periods
  previously measured in optical lines from the wind.}
  

   \maketitle
%

\section{Introduction}\label{introduction}

Herbig Ae/Be  stars, first defined by \citet{herbig60}, are young
intermediate-mass ($\approx 2-10~M_{\odot}$) stars predominantly located near star-forming
dark clouds. They show emission lines in their optical spectra, and their
placement in the HR diagram proves that they are pre-main sequence stars
\citep{strom72}. Herbig stars may therefore be considered to be 
intermediate-mass analogs of low-mass T Tauri stars (TTS), and in particular
of their accreting variant, the classical T Tau stars (CTTS). The analogy
between Herbig stars and CTTS extends to infrared excess emission indicative
of heated circumstellar material (e.g., disks) and photometric variability.
Herbig stars are important in the study of star formation because their
evolutionary scenario is intermediate between that of low-mass and high-mass
stars. In the former class, accretion phase, dispersal of the circumstellar disk, and
onset of hydrogen burning occur sequentially in time whereas in the latter, the stars 
enter their main-sequence phase while still being embedded and accreting, making 
their study much more difficult.
   
Like their main-sequence descendants, B and A-type stars, Herbig stars are
generally supposed to be radiative in their interiors. The lack of a thick 
convection zone would make the operation of a solar-like $\alpha-\omega$ 
dynamo impossible, and hence no magnetic fields are expected at the surface
of these stars except possibly fossil magnetic fields trapped in the star
since the initial cloud contraction phase. However, transient convection
may be present in some of these stars during a short phase of deuterium burning
in a shell, opening the possibility that some dynamo-generated, non-potential
fields develop \citep{palla93}; alternatively, a dynamo powered by rotational 
shear energy may generate some surface magnetic fields in rapidly rotating, 
accreting Herbig stars as well \citep{tout95}. Although difficult to detect, 
magnetic fields have recently indeed been measured on several Herbig stars,
with (longitudinal) field strengths up to a few 100~G \citep{donati97, hubrig04,wade05}.
In spectropolarimetry studies of 50 Herbig stars, \citet{wade05} detected magnetic fields
in 5 of them and discussed these stars to be progenitors of the magnetic Ap/Bp stars.
\citet{praderie86} have suggested magnetic activity in AB Aur based on their observations
of periodically variable  blue wings in the Mg\,{\sc ii} line of this star. 
Whatever the generation mechanism of magnetic fields, it may be important to include 
circumstellar disks in the picture as the magnetic fields may attach to the inner 
border of the disk so that they directly interact with accreting material.
Moreover, the presence of magnetic fields may also be important in the presence 
of stellar winds, because the wind could then be magnetically confined and
the plasma could be shock-heated to X-ray temperatures \citep[e.g.,][]{babel97}. 
We will address this point in Sect.~\ref{magn_wind}.

X-ray emission is among the best tracers of magnetic fields in stars, although it  
equally well diagnoses wind or accretion shocks.
\citet{damiani94} and
\citet{zinnecker94} were the first to systematically study X-ray emission from
Herbig stars with the {\it Einstein} X-ray observatory and ROSAT, respectively.
They reported surprisingly high detection rates of 11/31 and 11/21, respectively.
These (and subsequent) studies have investigated X-ray emission in the context of
other stellar parameters, with the following principal results: i) The X-ray luminosity,
$L_{\rm X}$, increases with the effective temperature ($T_{\rm eff}$) and the stellar
bolometric luminosity ($L_{*}$), although the ratio $L_{\rm X}/L_{*} \approx 
10^{-6}-10^{-5}$ is higher than in O stars ($L_{\rm X}/L_{*} \approx 10^{-7}$) but
lower than in T Tau stars ($L_{\rm X}/L_{*} \approx 10^{-4}-10^{-3}$). ii) $L_{\rm X}$
is not correlated with the projected equatorial velocity $v\sin i$ \citep{damiani94, zinnecker94}, 
thus pointing either to a magnetic dynamo saturation effect or to X-rays not related 
to dynamo-driven magnetic fields. iii) $L_{\rm X}$ does correlate with indicators of disks, 
accretion, and outflow such as infrared excess, mass accretion rate $\dot{M}$, and wind 
velocity or wind momentum flux. The latter finding may indeed support a non-magnetic 
origin of the X-rays, for example shocks in unstable winds analogous to more massive 
O stars \citep{damiani94, zinnecker94}.

On the other hand, variability, flares, and extremely high electron
temperatures $> 10$~MK clearly favor magnetic processes \citep{hamaguchi00,
hamaguchi05, skinner04, giardino04, stelzer06a} whether based on dynamo-generated or fossil
magnetic fields near the star, or star-disk magnetic fields.  Surface convection 
plays an important role in transferring energy into magnetic fields by stirring the 
magnetic footpoints, a process that leads to coronal heating and mass ejections on the 
Sun and in magnetically active stars. If the convection zones are shallow as in late-A 
or early F-type main-sequence stars, then  the magnetic dynamos appear to operate rather 
inefficiently, leading to modest X-ray luminosities of coronal sources that reveal very 
soft spectra \citep{panzera99}. If convection is absent - as in main-sequence A-type stars - 
then fossil magnetic fields are unlikely to build up non-potential configurations although 
magnetic activity is widespread among chemically peculiar Bp/Ap stars \citep{drake87, drake94}. 
Alternatively, however, winding-up magnetic fields connecting the star with the inner 
circumstellar disk may episodically release energy through reconnection, thus heating 
plasma and possibly ejecting plasmoids that contribute to jets often seen in young stars 
\citep{hayashi96, montmerle00, hamaguchi05}.



Alternatively, X-rays have been suggested to be formed in accretion shocks in
the CTTS TW Hya based on its exceptionally soft X-ray spectrum, 
indications for high electron densities ($\approx 10^{13}$ cm$^{-3}$), and 
anomalous abundances of N and Ne. \citet{swartz05} proposed
a similar scenario for the Herbig star HD 163296 based on its 
unusual soft spectrum (kT $\approx$ 0.5 keV).

A caveat in general is that the majority of Herbig stars are binaries or multiples \citep{feigelson03}. 
A close, unidentified T Tau companion that hides in the strong optical light could easily 
produce the observed X-rays  because X-ray luminosities and electron temperatures of Herbig 
stars are often quite similar to those of  T Tau stars (see for example \citealt{stelzer06a}). This emission is
commonly interpreted as solar-type coronal magnetic activity.  The binary hypothesis
has become a relevant model for flaring X-rays in at least two Herbig 
stars (MWC 297, \citealt{hamaguchi00}; field stars discussed as alternative 
X-ray sources and companion discovered by \citealt{vink05}; and
V892 Tau, \citealt{giardino04}; an 1.5-2$M_{\odot}$ companion was discovered by 
\citealt{smith05}). \citet{stelzer06a} studied a sample of 17 Herbig stars
observed with {\it Chandra} and detected 13 of them in X-rays. Of these 13 stars, 7
have a known visual or spectroscopic companion that could be the source of the
observed X-rays. Only 35\% of the detected sources cannot be explained by known companions.
Furthermore, in the latter work, X-ray properties of Herbig stars are found
to be very similar to X-rays properties of CTTS, leading to two possibilities: either 
the mechanism of the X-ray generation is similar for the two types of stars,
or the X-rays are generated by (partly unknown) low-mass companions. 

We report here the first high-resolution X-ray spectrum of a Herbig 
star, namely AB Aurigae, obtained with the Reflection Grating 
Spectrometer (RGS) on board {\it XMM-Newton}. Only the high 
resolution spectrum gives access to He-like line triplets, which
yield information on densities or UV radiation fields in X-ray 
emitting regions. This, in turn, constrains the source location
and extent \citep{behar04}. Further, high resolution spectroscopy
enables  accurate abundance determination using individual spectral 
lines (including lines of N and O forming at low temperatures and 
located at long wavelengths), and, together with CCD spectra 
obtained with the European Photon Imaging Cameras (EPIC), 
accurate thermal modeling.
Our analysis shows that AB Aur is another very soft source with a
moderate X-ray luminosity, but there are no signs of increased electron
densities nor elevated abundances. We will use the nearby, co-eval,
classical T Tau star SU Aur as an ideal comparison star to pinpoint fundamental
differences in these X-ray sources with largely differing interiors.

The structure of our paper is as follows. First, we describe our target in Sect.~\ref{target}.
We then present our data reduction procedures in Sect.~\ref{observations}. Sect.~\ref{results} presents our
results  from the X-ray spectroscopy, and Sect.~\ref{triplet} specifically analyzes
information from the He-like triplet of O\,{\sc vii}.
We discuss possible models in Sect.~\ref{discussion}, and conclude in Sect.~\ref{conclusions}.

\section{AB Aur}\label{target}

Table~\ref{tab2} summarizes the basic properties and the principal X-ray parameters
of AB Aur. For comparison, the properties of the Herbig star HD~163296, that has been
reported to reveal a soft X-ray spectrum \citep{swartz05}, and parameters of the 
CTTS SU Aur are also reported (see also Sect.~\ref{results} \& Sect.~\ref{discussion}).

\begin{table}[t!]
\begin{minipage}{0.48\textwidth}
\caption{Parameters for AB Aur, SU Aur, and HD~163296  \label{tab2} }
\begin{tabular}{lrrr}
\hline
\hline
Parameter     & AB Aur        & HD~163296       & SU Aur   \\  
\hline
Magnetic field    &  Y? (1)       & Y? (17)             & Y        \\
Spectrum      &B9.5e-A0 (2,3) & A1Ve (4)        & G2 (5)  \\
$L_{*}$ [$L_{\odot}]$&  49 (16)       & 30 (4)         & 9.9 (6) \\
Mass [$M_{\odot}]$& 2.7 (16)  & 2.3 (4)         & 1.9 (X) \\
Radius [$R_{\odot}]$& 2.3 (16) & 2.1 (4)         & 3.1 (X)  \\
$A_V$ [mag]   &  0.25 (23)     & 0.25 (4)        & 0.5 (7) \\
              &  0.47 (16)     &         &  \\
$T_{\rm eff}$ [K] &  9750 (8)     & 9300 (4)        & 5860 (6) \\
              &  10050 (16)     &         &  \\
Age [ Myr]    &  4 (16)       & 4  (9)          & 4 (7)   \\
Disk          &  Y (21)       & 450~AU (9)      & Y        \\
Companions?   & evidence (20)      & not det. (9)    & ...      \\
Radio spectrum & wind (11)     & wind (12)       & ...      \\
EW(H$\alpha$) [\AA] &  27 (13)      &                 & 2--6 (X)  \\
vsini [km~s$^{-1}$]       &  80 (18)       &  120 (19)       & 65 (22)  \\
$kT$ [keV]    &  0.46   (25)  & 0.49   (9)      & 1.9 (25)    \\
$\log L_{\rm X}$ [erg~s$^{-1}$]   &  $<$30.3 (3)  & 29.6 (9)        & 30.9 (25)  \\
              &  29.5 (10)    &                 &          \\
              &  29.6 (25)    &                 &          \\
$\log L_{\rm X}/L_{*}$&-5.8 (10)& -5.48 (9)     & -3.7 (25)  \\
              &  -5.6 (25)    &                 &          \\
$\log N_H$ [cm$^{-2}$]   &  20.7 (24, X) & 20.88 (9)       & 21.4 (25)   \\
X-ray variable? & slow (25)   & marginal? (9)   & flares (24)      \\
\hline        
\end{tabular}
\footnotetext{
REFERENCES given in parentheses: 
1     \citet{praderie86} - see text for further discussion; magnetic field assumed for CTTS.
2     \citet{hamann92};
3     \citet{hamaguchi05} (and references therein);
4      \citet{vandenancker98}; 
5      \citet{kenyon95}; 
6      \citet{luhman04}; 
7      \citet{dewarf03}; 
8      \citet{acke04}; 
9     \citet{swartz05} (and references therein);
10    \citet{zinnecker94};
11    \citet{skinner93};
12    \citet{blondel93}
13    \citet{herbig88};
14    \citet{grady00};
15    \citet{devine00};
16     deWarf \& Fitzpatrick, private communication; 
17     \citet{deleuil05};
18      \citet{boehm93};
19     \citet{hillenbrand92};
20     \citet{baines06};
21     \citet{corder05};
22     \citet{hartmann89};
23     \citet{roberge01};
24     \citet{robrade06};
25     this work.
X      see XEST survey results and tables in \citet{guedel06} for derivations and references. 
}
\end{minipage}
\end{table}

New, preliminary estimates of the surface temperature ({\it T}$_{\rm eff}$)
and stellar luminosity ({\it L}$_{*}$) of AB Aur have been derived
using the method developed by \citet{fitzpatrick99}.  By utilizing
the available UV (IUE: SWP + LWP) through optical (photometric: UBV) data,
the energy distribution can be modeled with surface fluxes represented
by R.~L. Kurucz's ATLAS13 atmospheric models \citep{kurucz93}.  See \citet{dewarf03} 
for a more detailed description of this procedure and
how it was implemented for SU Aur.  A complete
description of this analysis, as it pertains to AB Aur, is in preparation
(DeWarf et al., in preparation).

The rotation period of AB Aur is controversial. Recent observations 
suggest an inclination angle $i \sim 21.5^{\circ}$ (\citealt{corder05} and
references therein), that would suggest, using   $v$sin$i$ and radius 
from Table~\ref{tab2}, a rotation period $P=12.9$~hr.
For an extreme value of 
$i = 70^{\circ}$ previously reported for the disk inclination 
(see references in \citealt{corder05}), we obtain $P=33$~hr.
A period of this order is supported by modulations in Mg\,{\sc ii}
and Ca\,{\sc ii} lines \citep{praderie86, catala86} 
and in photospheric lines \citep{catala99} that reveal periods $P=32-34$~hr
and $P=43-45$~hr (see Sect.~\ref{atm} for a detailed discussion).


To our knowledge, no direct detection of surface magnetic fields
of AB Aur has so far been obtained. \citet{catala99} estimated an upper limit
to the strength of the magnetic field of 300 G. 

\section{Observations and data analysis}\label{observations}

AB Aur was detected in an {\it XMM-Newton} \citep{jansen01} observation pointing at the nearby 
CTTS SU Aur (separation between SU Aur and AB Aur: $\approx 2.5\arcmin$). 
The observation was retrieved from the archive as part of the {\it XMM-Newton Extended Survey
of Taurus} (XEST) described in \citet{guedel06} (XEST observation number
of AB Aur: XEST-26-043, and of SU Aur: XEST-26-067). Table~\ref{obslog}
summarizes observing parameters.
The two European Photon Imaging Cameras  (EPICs) of the MOS type \citep{turner01} 
and the Reflection Grating Spectrometers (RGSs; \citealt{denherder01}) were active 
during the observation, while the EPIC PN camera was out of operation.
Both MOS instruments observed in full frame mode and used the thick filter to
suppress excessive optical load from AB Aur.  
The EPIC detectors operate in the energy range of 0.15--15.0 keV with a medium 
spectral resolution of approximately $E/ \Delta E = 20-50$.  
The RGSs are suited for high-resolution spectroscopy in the wavelength
range of 6--35~\AA\ and have a resolution of $\Delta \lambda \approx 60-76~$m\AA. 
The RGS detectors contained the dispersed spectra of both AB Aur and SU Aur with
sufficient separation to make mutual spectral contamination negligible (see below). 
The spectrum of SU Aur has recently been discussed by \citet{robrade06}.
\begin{table}[t]
\caption{Observing log\label{obslog} }
\begin{tabular}{lrrr}
\hline
\hline
XMM-Newton ObsID         & 0101440801         \\
Boresight RA (J2000.0)       & 4$^{\rm h}$ 55$^{\rm h}$ 59.0$^{\rm s}$         \\
Boresight $\delta$ (J2000.0) & 30$^{\circ}$    34$\arcmin$  02$\arcsec$        \\
Start time  (UT)             & 2001-09-21\ 01:34:17  \\
Stop time   (UT)             & 2001-09-22\ 13:34:31  \\
Exposure time  (s)           & 129614   \\
MOS Mode and Filter          & Full Frame, thick filter    \\
\hline
\end{tabular}
\end{table}   

The data were reduced using the Science Analysis System (SAS) 
version 6.1. The EPIC MOS data were reduced using the task {\it emchain}
and the sources were detected using the maximum likelihood detection
algorithm {\it emldetect} (see \citealt{guedel06} for further details).
To extract the two spectra from each RGS detector, we proceeded as follows.
We first applied the standard processing performed by the RGS metatask 
{\it rgsproc} to each source position. We extracted the total 
(source+background) spectra and the background spectra separately.
For both SU Aur and AB Aur, we included 85\% of the
cross-dispersion Point Spread Function (PSF) as also done by  \citet{telleschi06} ({\it xpsfincl = 85}). 
For the background, the exclusion region with
regard to  the cross-dispersion PSF  and the inclusion region of the pulse-height distribution
were kept at default values (95\% of the PSF and 90\% of the 
pulse-height distribution, respectively, i.e., {\it xpsfexcl = 95} and {\it pdistincl = 90}).
We verified that this way the spatial extraction regions on the
detector were adjacent to each other but just not overlapping. This means
that each spectrum collects approximately (100-85)/2 = 7.5\% of the counts of the
other spectrum. The contamination is such that counts from SU Aur
are shifted in wavelength by approximately  -0.4~\AA\ in the AB Aur spectrum within
 the wavelength range of interest here (12--22~\AA), and
contaminating counts from AB Aur are shifted by +0.4~\AA\ in the SU Aur spectrum.

The background defined for each source outside its source extraction region
is now still contaminated by the other source. Therefore, we defined the 
coordinate of the secondary source (AB Aur for the SU Aur
spectrum and vice versa),  added them to the source list using the {\it rgssources}
task, and computed a new extraction map, with the secondary source excluded 
from the background, using the {\it rgsregion} task.
Finally, we extracted each spectrum again, using only the regions outside the SU
Aur and AB Aur source regions for the background spectra.
The background for the AB Aur spectrum is extracted outside the two adjacent
source regions. The background portion on the far side of
the SU Aur spectrum again comprises approximately 7.5\% of the SU Aur source 
counts. Because this background spectrum is subtracted from the total spectrum extracted
at the AB Aur position, the contamination is approximately corrected for.
Further considering the low S/N, the similar 
line fluxes in both spectra, and the line-dominated spectrum of AB Aur (see below,
while most of the counts in SU Aur are in continuum and therefore distributed in wavelength), 
the mutual contamination is negligible and much below the noise level. For example, 
no significant mutual contamination is seen by the strong O\,{\sc viii} lines above the 
noise level in either of the spectra (see presentation of the RGS spectra in 
Fig.~\ref{rgsspec} below).
The periods affected by high background flaring were excluded from the spectral
analysis.

Because we wanted to put weight on the high-resolution spectra obtained with RGS,
we performed the data analysis on both RGS spectra, adding only the short-wavelength portion
of one of the two MOS detectors in order to access line features of Mg, Si and Fe that 
are important for our abundance analysis (see \citealt{telleschi05}).  As in \citet{telleschi06}, we used
MOS1 for SU Aur for this purpose, while for AB Aur, we preferred MOS2 because an unidentified 
background feature distorted the MOS1 spectrum around a wavelength of 9.3~\AA\ (not present
in MOS2).  The wavelength intervals of each instrument
used for the data analysis are summarized in Table~\ref{tab:ranges}. As a check, we confirmed
our results by using the entire useful energy range of MOS (0.2--10~keV), and the
results are consistent with those reported here.

\begin{table}
\caption{Wavelength ranges used for the spectral fitting}             
\label{tab:ranges}      
\centering                          
\begin{tabular}{c c c}        
\hline\hline                 
Instrument & AB Aur & SU Aur \\
\hline                        
MOS1  & --              & 1.5--9.35 \AA  \\
MOS2  & 1.5--9.35 \AA   & --  \\
RGS1  & 10.0--28.0 \AA & 8.3--25.0 \AA  \\
RGS2  & 8.3--26.5 \AA  & 8.3--25.0 \AA  \\
\hline                                   
\end{tabular}
\end{table}

We fitted the spectra in XSPEC \citep{arnaud96} using optically-thin collisionally-ionized plasma
models calculated with the Astrophysical Plasma Emission Code (APEC; \citealt{smith01}).
In order to account for calibration discrepancies between the RGS and the MOS 
instruments, we added constants as effective-area renormalization factors.
These factors were  fixed at 1.0 for MOS and at 1.05 for both  RGS (see \citealt{kirsch04}).   

Since we used the $\chi^2$-statistic for our spectral fitting, we binned the
spectra to a minimum of 20 counts per bin for the RGS and a minimum of
15 counts per bin for the MOS. The resulting bin width varies between 0.04
\AA~(in the O\,{\sc vii} line) and 0.73 \AA~(just longward of 
the O\,{\sc viii} Ly$\alpha$ line) in the RGS,
and between 0.15 \AA~ and 1.4 \AA~in the MOS spectrum.

For AB Aur the hydrogen column density $N_H$ was fixed at the value found in
the XEST survey analysis, namely $5 \times 10^{20}$~cm$^{-2}$, which is consistent with
the recent measurements of $A_V = 0.25$~mag  reported by \citet{roberge01}, assuming
a standard conversion $N_H \approx 2.0\times 10^{21}~A_V$ applicable to
the interstellar medium (\citealt{vuong03}, and references therein). 
This value agrees quite well with the $N_H$ found explicitly
by \citet{roberge01}, $N_H = 4.4 \times 10^{20}$~cm$^{-2}$.  
For SU Aur, we needed to fit $N_H$ in order to get
a good fit to the RGS spectra.


The spectra were fitted with two different models: a model describing a differential emission 
measure distribution (EMD) approximated by two power laws as used in the XEST survey analysis \citep{guedel06}, 
and a model with two or three isothermal plasma components. For both models we applied a procedure
in which we simultaneously fitted large wavelength intervals of the three spectra with template
spectra computed in XSPEC. Alternative, iterative methods based on extracted line fluxes are
not feasible given the low S/N ratio of our spectra. There are only few explicitly measurable lines,
and each set of lines from a given element is confined to narrow formation temperature 
interval. Also, such methods cannot be applied to
MOS CCD data. Previous studies have shown excellent agreement between iterative methods
and the method applied here \citep{telleschi05}. 
The EMD model is approximated by a grid
of isothermal components spaced regularly by 0.1 dex in $\log T$ in such a way that the lower-$T$ and the
higher-$T$ portions are each described by a power law. This
model has been suggested from our previous work on high-resolution X-ray spectroscopy
of young solar analogs \citep{telleschi05}, and is described by 

\begin{equation}\label{eq:dem}
Q(T) = \left\{ \begin{array}{ll} EM_0 \cdot (T / T_0)^{\alpha}\quad   &, for ~ T \le  T_0 \\
                                      EM_0 \cdot (T / T_0)^{\beta}\quad     &, for ~ T >  T_0 \end{array} \right.
\end{equation}
where $T_0$ is the temperature where the two power-laws cross, and $EM_0$ is the 
EM per $\log T$ at this crossing point. The slopes of the power laws below and above
$T_0$ are $\alpha$ and $\beta$, respectively. The power laws have high-energy and low-energy cut-offs 
at $\log T = 8$ and $\log T = 6$, respectively. Given the poor energy resolution of 
CCDs below 1 keV, we had fixed $\alpha$ at a value of +2 throughout the XEST survey, and we 
will do so here for comparison reasons. A slope of 2 is compatible with findings from
DEM analyses in other pre-main sequence and main-sequence active stars \citep{argiroffi04,
telleschi05}.

However, because the X-ray emission of a Herbig Ae/Be star could be non-coronal, or the
coronal thermal structure could be unlike that in late-type T Tau stars, it is possible that 
the simplified EMD model proposed here is inappropriate. In a separate approach, we will
therefore also fit $\alpha$. Further, because the power-law EMD structure might be 
inappropriate altogether, we will test our results using a multi-temperature model
that can reveal the temperatures where most EM resides.
We therefore analyzed the AB Aur spectra with a 2-component model.
A third component was not needed, given the low signal-to-noise ratio of the
spectrum and, as we will describe below, its rather narrow range of temperatures in 
which emitting plasma is found.
Because spectral lines are formed over an interval of typically 
0.3 dex in temperature, a temperature range of 2-7 MK (the range
of formation temperatures present in the spectrum) will require no 
more than two thermal components.
We also analyzed the SU Aur spectra with a 3-component model.

For both approaches, we fitted the abundances of the lines observed in the spectra
simultaneously with the thermal models.
Abundances that do not show significant features in the spectra, such as 
those of C, S, Ar, Ca and Al (and N for SU Aur), were fixed at the values 
used in the general XEST survey data analysis (C=0.45, N=0.788, S= 0.417, 
Ar=0.55, Ca=0.195, and Al=0.5, \citealt{guedel06}). These abundances 
were thus arranged in such a way that they describe a weak ``inverse 
First Ionization Potential Effect'' (higher FIP implies higher abundances relative
to the photospheric values, the latter assumed to be solar), as often observed 
in young stars

Finally, we computed the X-ray luminosities, $L_{\rm X}$, from the spectral model in the energy range
of 0.3--10.0~keV, assuming a distance of 140~pc (\citealt{dewarf03} and references therein).

\section{Results}\label{results}

\subsection{Light curves}\label{sec:lc}

\begin{figure}
\centering
\includegraphics[angle=-0, width=0.48\textwidth]{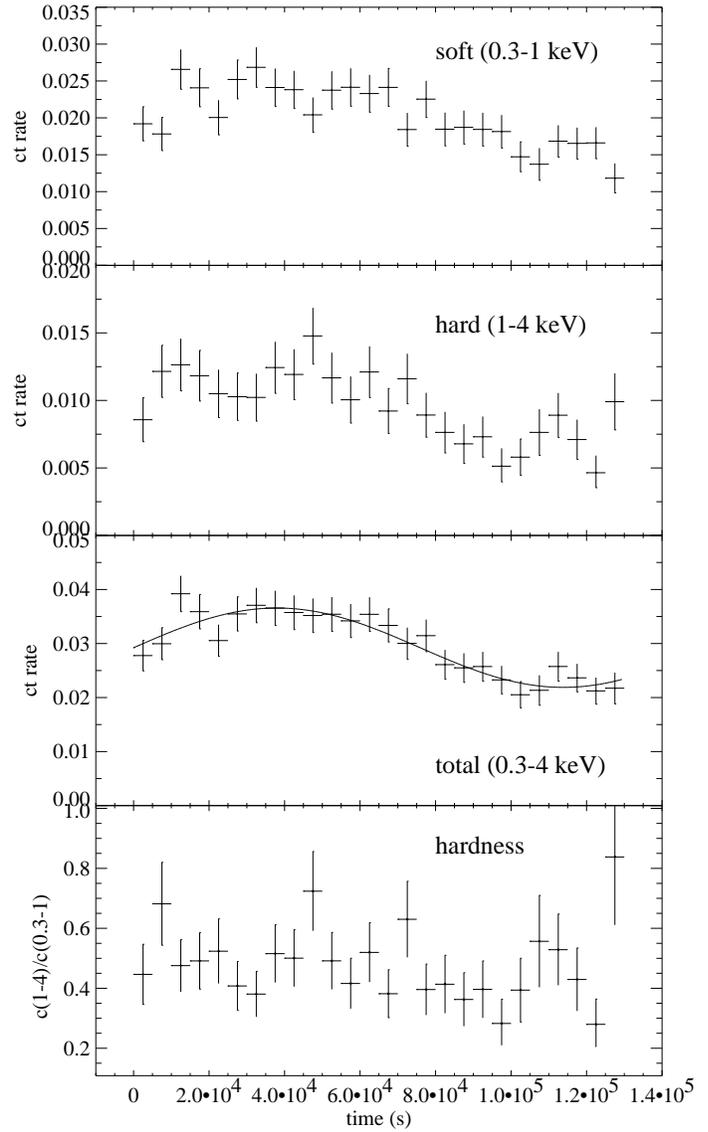}
\vskip -0.6truecm
\caption{Light curves of AB Aur, combining counts from both MOS detectors, binned to 5 ks. Panels from 
     top to bottom: Soft counts in the 0.3--1~keV range; hard counts in the 1-4~keV range;
     counts in the combined range, 0.3-4~keV; hardness, defined by the ratio of hard to
     soft counts.}
      \label{lc}
\end{figure}

We first present the X-ray light curve of AB Aur in Fig.~\ref{lc}.
We have co-added counts from both EPIC MOS detectors after
background subtraction. The background contribution to
the total light curve was in fact negligible except during the final
5 ks when a background flare occurred that reached $\approx 50$~\%
of the total count rate. The observed variability can therefore
not be attributed to imperfect background treatment but is
intrinsic to the stellar source. We see slow variability on time scales of about one day,
by somewhat less than a factor of two. 
In the variability analysis of the XEST sources presented
by \citet{stelzer06b}, the variablity of AB Aur was confirmed by
the Kolmogorov-Smirnov test. No significant trend is seen
in the hardness, defined as the ratio between the count rates in the
hard (1-4~keV) and the soft (0.3-1~keV) band. 
We have fitted the light curve with a sine function which is also plotted 
in Fig.~\ref{lc}. The fit is excellent ($\chi^2 = 12.9$ for 22 dof) 
with a period of $42.2^{+4.4}_{-3.7}$~hr in the total band (in the soft and hard band 
we find $48.5^{+11.6}_{-7.8}$~hr and $40.4^{+5.2}_{-4.2}$ hr, respectively). A similar modulation of 
$\approx 42-45$~hr was found previously for the He\,{\sc i} and Mg\,{\sc ii} lines (see discussion below).

We have studied the light curve variability with a statistical test,
to assess the presence of a flare in the last 30 ks, where an increase in
the count rates can be seen, particularly in the hard and total light curves
(second and third panel in Fig.~\ref{lc}, respectively).
We first tested our data against a constant count-rate model using the
$\chi^2$-statistic. We find the probability for this part of the 
light curve to be constant, $P$(const)=0.73 for the total band, and $P$(const)=0.30
for the hard band. 
As a second test, we used the Kolmogorov-Smirnov statistic, to obtain $P$(const)=0.09 
and $P$(const)=0.07 for the total and hard light curves, respectively.
The variability of the last 30 ks in the light curves is thus at best marginal.

\subsection{Spectra}

\begin{figure}
\centering
\includegraphics[angle=-90, width= 0.48 \textwidth]{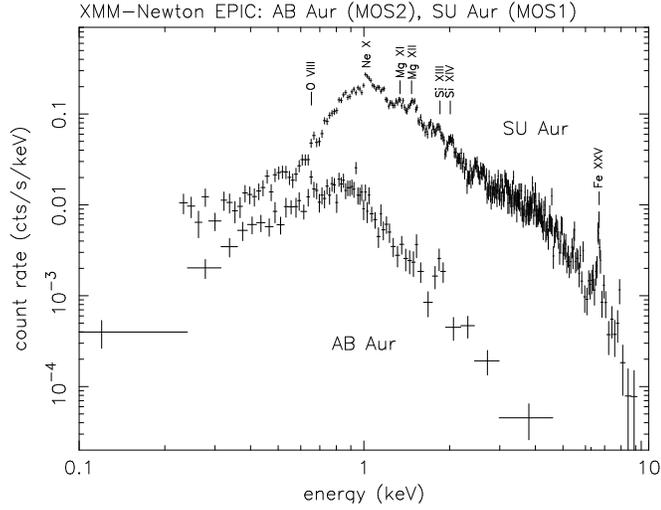}
  \caption{EPIC spectra of SU Aur (MOS1) and AB Aur (MOS2). Important lines are labeled.}
      \label{epicspec}
\end{figure}

Fig.~\ref{epicspec}  compares the background-subtracted
EPIC MOS spectra of AB Aur and SU Aur. A number
of spectral differences are obvious. The spectrum of
SU Aur reveals signatures of a very hot coronal plasma, with outstanding
lines of Mg\,{\sc xi}, Mg\,{\sc xii}, Si\,{\sc xiii} and, most notably, the Fe~K complex
mostly due to Fe\,{\sc xxv} at 6.7~keV. A steep drop toward the lowest
energies indicates considerable photoelectric absorption. In contrast,
the AB Aur spectrum falls off rapidly above 1~keV, showing its peak
flux around 0.7-0.9~keV. This flux peak is mainly due to lines
of Fe\,{\sc xvii}. These spectral properties let us anticipate a
soft source for AB Aur, and the shallow fall-off toward low energies indicates
rather low photoelectric absorption.

\begin{figure*}
\centering
\hbox{
\hskip -0.5truecm\includegraphics[width=1.0\textwidth]{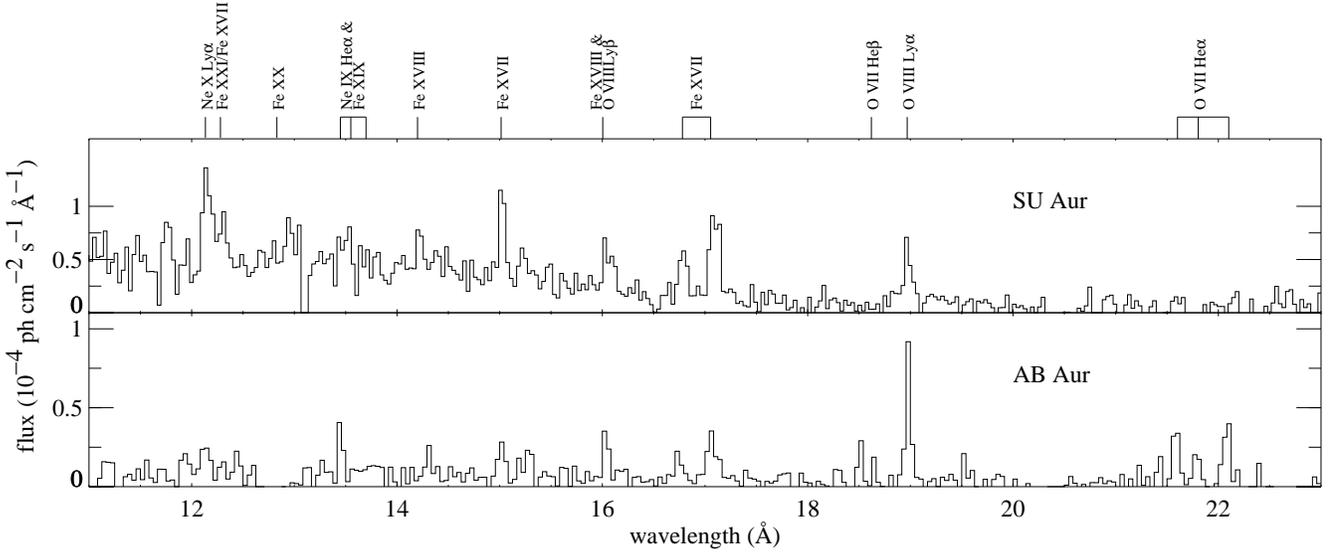}
}
\caption{Fluxed, coadded RGS1+RGS2 spectra of SU Aur (top) and AB Aur (bottom). 
The spectra are background subtracted.}
      \label{rgsspec}
\end{figure*}

Fig.~\ref{rgsspec} shows fluxed, co-added RGS1+2 spectra of SU Aur (top)
and AB Aur (bottom) in the line dominated region. 
The spectra have been rebinned to a bin width of 0.042 \AA~for AB Aur and
0.035 \AA~for SU Aur. These spectra further corroborate the
differences between the two X-ray sources. Whereas SU Aur reveals
a strong continuum, indicating a hot source, there is little evidence for
continuum emission in AB Aur. Further, the flux ratio between the
Ne\,{\sc x} Ly$\alpha$ line at 12.1~\AA\  (formation temperature 6.3 MK) 
and the Ne\,{\sc ix} resonance line at 13.44~\AA\  (formation temperature 4 MK)
is considerably higher in SU Aur than in AB Aur, again emphasizing the
dominance of hot plasma in the former. Furthermore, the Ne\,{\sc ix} line feature
at 13.5~\AA~in SU Aur is dominated by Fe\,{\sc xix}, 
which is formed at higher temperatures (formation temperature 7.9 MK). 
Two further features are
striking: First, the Fe lines of SU Aur are very strong,
dominating the spectrum and comparing in flux with the O\,{\sc viii} Ly$\alpha$
and the Ne\,{\sc x} Ly$\alpha$ lines (although the O\,{\sc viii} line is partly suppressed
by photoelectric absorption). Such line ratios are unusual among
very active, main-sequence solar analogs where Fe line fluxes are modest
due to low Fe abundance \citep{telleschi05}. We thus anticipate an
unusually high abundance of Fe in this spectrum.

The second feature of interest here is the unusually high flux of  the
O\,{\sc vii} lines of AB Aur, despite some photoelectric absorption that suppresses the
flux at these wavelengths. The total O\,{\sc vii} flux appears to be similar to the flux in
the O\,{\sc viii} Ly$\alpha$ line, which is a property of very inactive
stellar coronae with  temperatures of 3--5~MK \citep{telleschi05}.
(Note that the feature at 18.5~\AA\ is neither  coincident with the O\,{\sc vii} He$\beta$
line nor due to contamination by the O\,{\sc viii} Ly$\alpha$ line of SU Aur. Those two
line features would both be slightly but significantly longward of this wavelength, namely at 18.6 \AA. 
The 18.5~\AA\ feature is due to a 3$\sigma$ spike exclusively in the RGS2 detector 
and is therefore spurious). We note that the O\,{\sc vii} lines are not present in the 
spectrum of SU Aur due to considerable photoelectric absorption.

\subsection{Thermal structure}

\begin{table*}
\caption{Results of spectral model interpretation$^1$}             
\label{tab:fitres}      
\centering                          
\begin{tabular}{l c c c c c}        
\hline\hline                 
   &  \multicolumn{3}{c}{AB Aur} & \multicolumn{2}{c}{SU Aur} \\
  & \multicolumn{3}{c}{\hrulefill} & \multicolumn{2}{c}{\hrulefill} \\
  & 2T model & DEM model~$^2$ & DEM model~$^3$ & 3T model & DEM model~$^2$  \\
\hline                        
$N_H$ [$10^{21}$ cm$^{-2}]$ & = 0.5~$^2$  & = 0.5~$^4$ & = 0.5~$^4$&  3.1 (3.0, 3.3) &  3.2 (3.0, 3.4)\\
\hline
N~$^5$    & 0.45 (0.22, 0.78)   & 0.57 (0.30, 1.13)  & 0.58 (0.27, 1.13)  & --       & --\\
O~$^5$    & 0.20 (0.16, 0.42)   & 0.22 (0.13, 0.32)  & 0.27 (0.17, 0.45) &  0.50 (0.40, 0.61)& 0.30 (0.24, 0.38)\\
Ne~$^5$   & 0.60 (0.41, 0.88)   & 0.62 (0.48, 1.04)  & 0.77 (0.49, 1.22) & 0.61 (0.46, 0.78)& 0.38 (0.28, 0.53)\\
Mg~$^5$   & 0.21 (0.09, 0.38)   & 0.28 (0.13, 0.74)  & 0.34 (0.14, 0.62) & 1.51 (1.33, 1.74)& 1.17 (1.05, 1.43)\\
Si~$^5$   & 0.70 (0.48, 1.01)   & 0.90 (0.60, 1.32)  & 0.96 (0.64, 1.51) & 0.79 (0.69, 0.92)& 0.64 (0.57, 0.79)\\
S~$^5$    & --                  & --                 & --        & 0.64 (0.49, 0.79)& 0.57 (0.45, 0.72)\\
Fe~$^5$  & 0.23 (0.17, 0.32)   & 0.29 (0.22, 0.47)  & 0.37 (0.24, 0.61)  & 0.81 (0.73, 0.91)& 0.67 (0.61, 0.77)\\
\hline
$T_1$ [MK] & 2.45 (2.10, 2.81)  &  -- &  -- & 7.54 (7.30, 7.75)  &  --\\
$T_2$ [MK] & 6.99 (6.62, 7.41)  &  -- &  -- & 13.97 (12.05, 18.06)   & --\\
$T_3$ [MK] & --                &  -- &  -- & 38.14 (35.68, 41.14)   & --\\
EM$_1$ [$10^{52}$ cm$^{-3}$]  & 2.11 (0.87, 3.67)  &  -- &  -- & 11.27 (9.18, 14.04) & --\\
EM$_2$ [$10^{52}$ cm$^{-3}$] & 3.44 (3.06, 4.31)  &  -- &  -- & 7.25 (4.53, 14.35)  & --\\
EM$_3$ [$10^{52}$ cm$^{-3}$] & --                &  -- &  -- & 30.49 (24.66, 33.21)& --\\
\hline
$T_0$  [MK]   & --   & 4.38 (2.69, 5.68) & 6.33 (5.26, 8.32) & --       & 7.66 (6.95, 8.13)\\
$\alpha$   & --   & = 2.00  & 0.47 (0.00, 1.20)  & --    & = 2.00\\
$\beta$   & --   & -1.9 (-2.57, -1.52)  & -2.05 (-3.00, -1.57)  & --    & -0.05 (-0.11, 0.06)\\
EM  [$10^{52}$ cm$^{-3}$]$^6$    & --   & 5.21   & 5.08   & --  &  57.84\\
\hline
$T_{\rm av}$  [MK]  &  4.69  & 4.71  & 4.07  & 22.65 &  20.07 \\
\hline
$L_{\rm X}$   [$10^{30}$ erg/s]$^7$ &  0.40  & 0.39  & 0.39  & 7.40  &  7.79 \\
$L_{\rm X}$   [$10^{30}$ erg/s]$^8$ &  0.55   & 0.53  & 0.56  & 8.53  & 9.15  \\
\hline
$\chi^2_{\rm red}$   & 1.06  & 1.02  & 1.00  & 1.20 & 1.23\\
dof        & 78    &   79   &   78   & 443 & 446\\
\hline                                   
\multicolumn{6}{l}{$^1$ 68\% error ranges are given in parentheses.}\\
\multicolumn{6}{l}{$^2$ Fitted with $\alpha$ held fixed at 2 like in the XEST survey \citep{guedel06}.}\\
\multicolumn{6}{l}{$^3$ Fitted with $\alpha$ being a free parameter.}\\
\multicolumn{6}{l}{$^4$ Held fixed at value found in the XEST survey \citep{guedel06}.}\\
\multicolumn{6}{l}{$^5$ Element abundances are with respect to solar values given by \citet{anders89} (\citealt{grevesse99} for Fe).}\\
\multicolumn{6}{l}{$^6$ Total EM integrated over temperature bins between $\log T=6-7.9$ [K] (see \citealt{guedel06} for more details).}\\
\multicolumn{6}{l}{$^7$ Determined in the 0.3-10.0 keV band.}\\
\multicolumn{6}{l}{$^8$ Determined in the 0.1-10.0 keV band.}\\
\end{tabular}
\end{table*}

\begin{figure*}
\centering
\hbox{
\hskip -0.5truecm\includegraphics[width=1.0\textwidth]{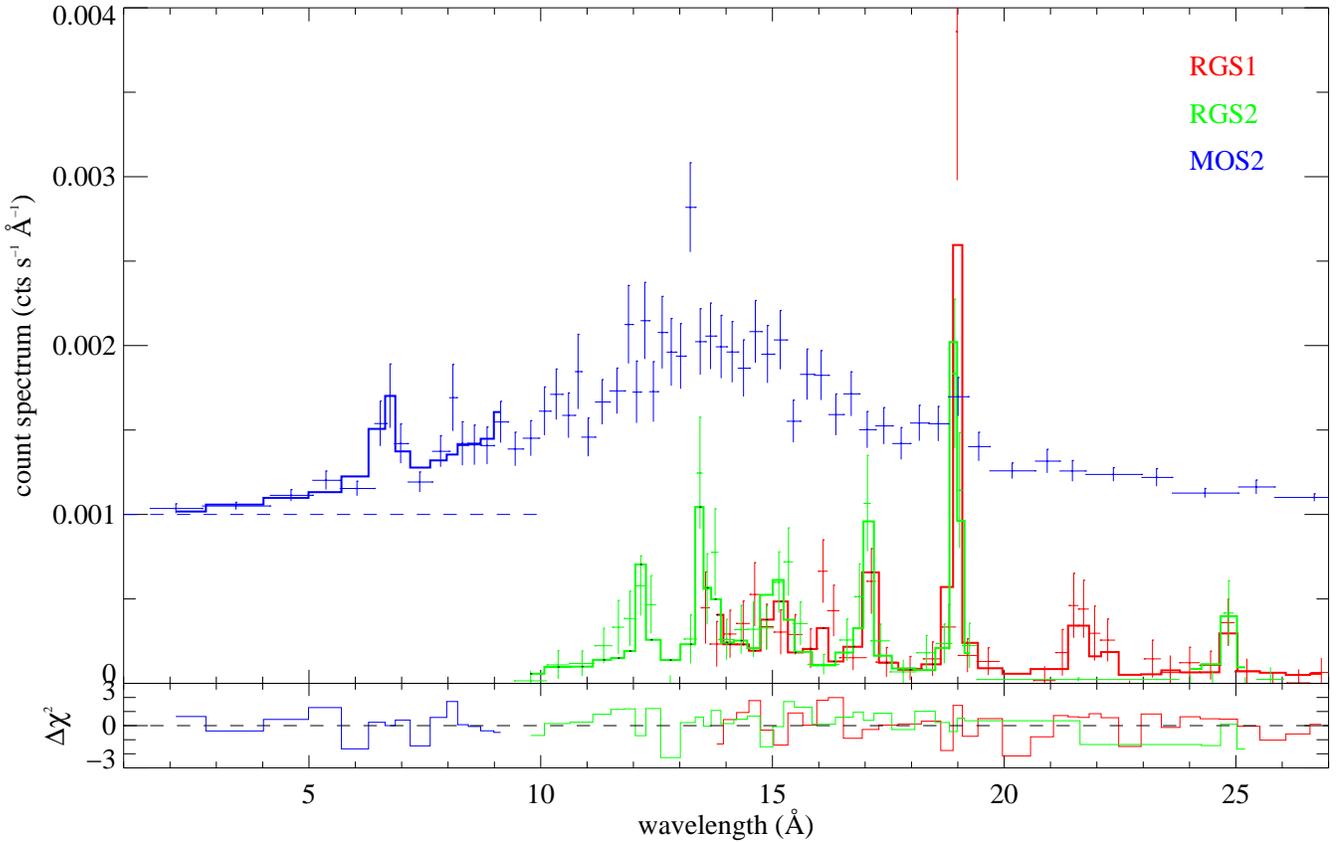}
}
\caption{Data and fitted spectrum of AB Aur (EMD model). The best-fit model is shown by the histograms
         in the wavelength region used for the fit. For plotting purposes, the MOS2 spectrum
         has been shifted along the y-axis by 0.001 cts s$^{-1}$ \AA$^{-1}$.}
      \label{fit}
\end{figure*}

We now present the numerical results of our spectral fits to these data.
Table~\ref{tab:fitres} lists the numerical results for both models
and both stars. In Figure~\ref{fit} we plot the data together 
with the best fit of the EMD model for AB Aur.
\footnote{The low bin at 16 \AA~ in RGS2, coincident with the 
O\,{\sc viii}+Fe\,{\sc xviii} lines, corresponds to a CCD gap.
At the same time, somewhat increased flux in the RGS1 spectrum slightly
longward of the two lines, probably due to imperfect background
subtraction, increases the RGS1 flux in the respective bin in 
Fig.~\ref{fit} above our fit. We checked at higher resolution that the 
O\,{\sc viii}+Fe\,{\sc xviii} line is indeed correctly fitted.}
We define the average coronal temperature, $T_{\rm av}$, as the
logarithmic average of all temperatures used in the model, applying
the corresponding EM as weights. This measure corresponds to the electron
temperature itself for an isothermal plasma. For a continuous
EMD as discussed here, $T_{\rm av}$ represents a temperature grossly
characteristic of the spectral shape. This is also true for multi-component
plasma (e.g., 2-$T$ or 3-$T$ plasmas) although there might be no plasma present 
at $T_{\rm av}$ itself. This also occurs when 1-$T$ fits are made to low-quality 
coronal spectra.

Both  the multi-temperature and the EMD models provide similar 
results: $T_{\rm av}$ and the X-ray luminosities are similar.
The 2$T$ or 3$T$-model fits 
and the EMD model fits show similar $\chi^2$ values, and are therefore
of statistically equal quality.
Further, most of the abundance values agree favorably
between the two methods, confirming the robustness of our results.

For AB Aur we report the results for EMD models with $\alpha$ frozen
at 2 and with $\alpha$ free (second and third columns of Table~\ref{tab:fitres},
respectively). The
peak temperature $T_{0}$ increases from 4.4 MK for $\alpha$ fixed at 2 to 
6.3 MK for free $\alpha$. However, the main results are in very close agreement for 
the two approaches: the abundances are consistent within the error bars, the X-ray 
luminosities are essentially unchanged, and $T_{\rm av}$ is only slightly smaller when we 
fit $\alpha$. This suggests that the two solutions are equivalent. 
On the other hand, the errors are larger if $\alpha$ is a free fit parameter 
(except for the Mg abundance). Therefore, 
and for consistency with the fits of the EPIC spectra in the XEST 
survey, we henceforth use the results from the EMD fit with $\alpha$ fixed
at a value of 2.

The $N_H$ value for SU Aur was found to be  
$ N_H = (3.1-3.2) \times 10^{21}$~cm$^{-2}$, which agrees well with the result
reported by  \citet{skinner98} from an ASCA observation,  $ N_H = 2.8 \times 10^{21}$~cm$^{-2}$ 
and is only somewhat higher than expected from  $A_V = 0.9$~mag \citep{kenyon95}.

The results confirm the peculiar thermal structure of the AB Aur source.
The average temperature from the 2-$T$ fit amounts to 4.7~MK, while
the EMD model peaks at 4.4~MK 
and rapidly falls off toward higher temperatures, with a power-law slope of -1.9. 
Such  low temperatures are unusual for  coronae of young stars where usually
temperatures in excess of 10~MK are found. An example is SU Aur:
The 3-$T$ model shows the largest amount of EM at the highest temperature 
at $\approx$ 40 MK, with   $T_{\rm av} \approx 23$~MK. Again, the EMD 
model supports this finding, where we find $T_0$ at 7.7~MK beyond which 
the EMD is nearly flat, resulting in an average temperature of 20~MK.
Franciosini et al. (2006, in preparation) have analyzed the same RGS spectrum of
SU Aur using a line-based analysis to derive the EMD from the measurement
of individual line fluxes. Their results are consistent with ours: the EMD peaks
at $T=10$ MK, with an indication of a significant amount of material above
$\sim$20 MK; however, below the peak they find a steeper slope ($\sim$ 3) than
adopted here.

\subsection{Abundances}

\begin{figure*}
\centering
   \centering
   \centerline{\hbox{
\includegraphics[width=0.5 \textwidth]{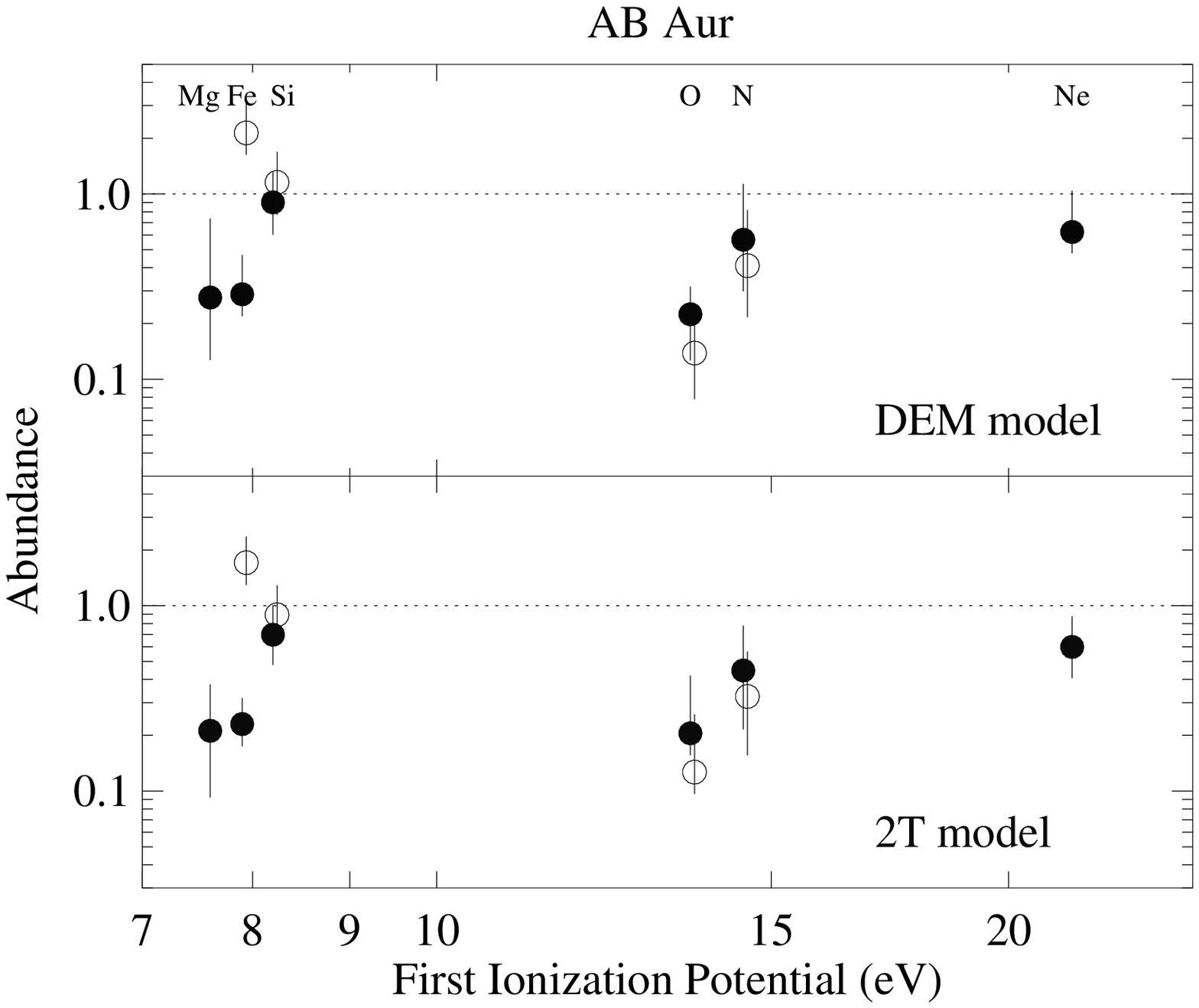}
\includegraphics[width=0.5 \textwidth]{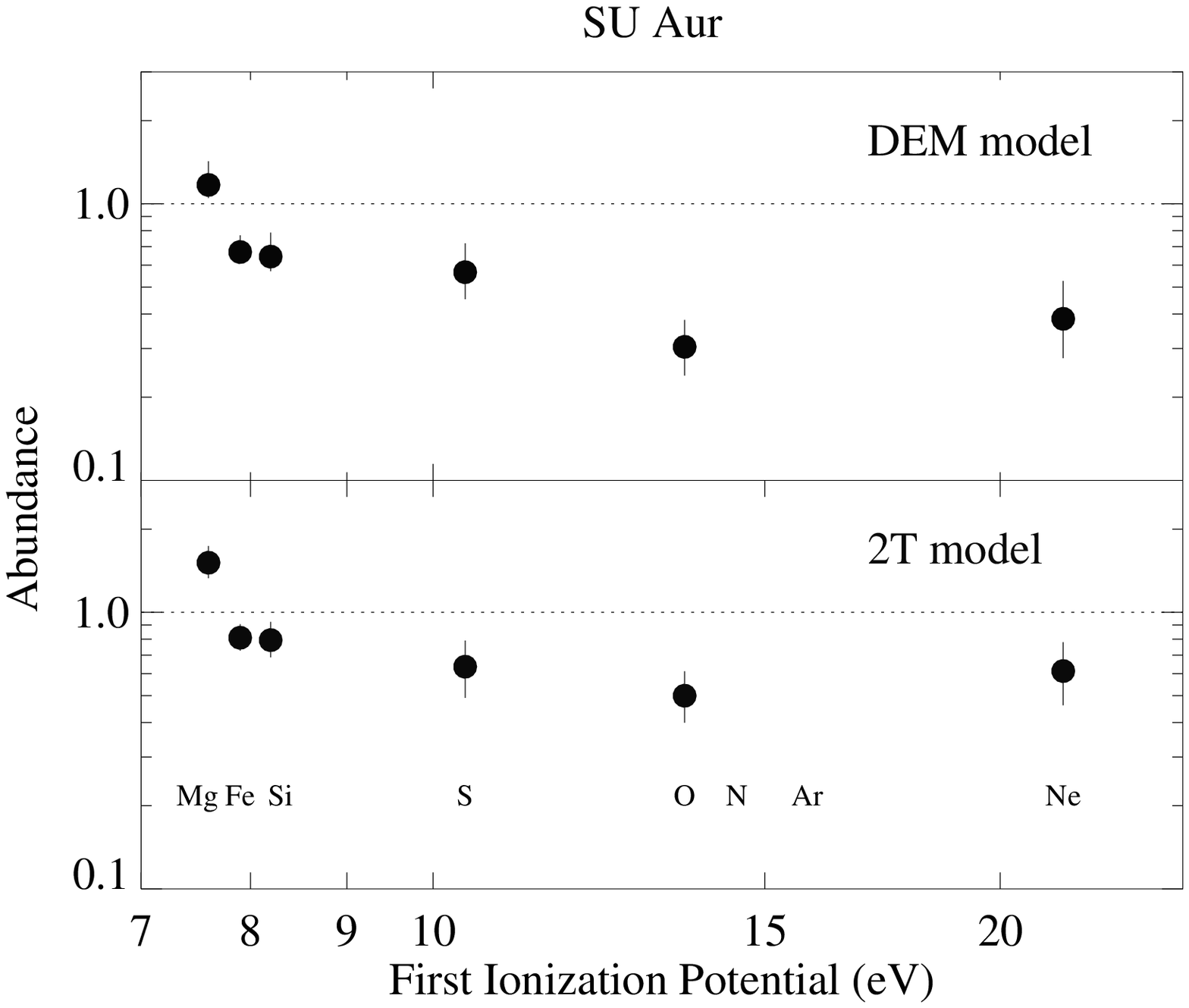}
    }}
  \caption{Abundances of AB Aur and SU Aur as a function of FIP. 
  Filled circles: normalized to solar photospheric ratios 
  (\citealt{anders89}; for Fe, \citealt{grevesse99}); open circles: normalized to AB Aur 
  photospheric values \citep{acke04}.}
      \label{abun}
\end{figure*}

The element abundances in the X-ray sources are
plotted in Fig.~\ref{abun}. The filled circles  
designate abundance ratios relative to the solar photospheric mix  (\citealt{anders89}; \citealt{grevesse99} for Fe).
The abundances of AB Aur show a nearly flat distribution; neither a strong First Ionization Potential
(FIP) effect (i.e., overabundant low-FIP elements) nor a strong inverse FIP effect (i.e., increasing 
abundances with increasing FIP) is seen, in contrast to the usual findings in young, active
stars (e.g., \citealt{argiroffi04, telleschi06}). Because the emitting material originates, if located
in a corona or a stellar wind, in the stellar photosphere, a comparison of the X-ray derived
abundances with photospheric values will be important. Fortunately, a few photospheric abundances
have recently been measured for AB Aur \citep{acke04}, and the resulting normalized 
abundance ratios are shown by the open circles\footnote{The photospheric abundances of AB Aur
are, with respect to the solar photosphere (error ranges in parentheses): 
N: 1.38; O: 1.62 (1.38-1.91); Si: 0.78 (0.72-0.83);
Fe: 0.13 (0.10-0.18), referring to the solar abundances of \citet{anders89} for N, O, and Si,
and those of \citet{grevesse99} for Fe.}.  \citet{acke04} report a particularly low 
photospheric abundance of Fe ($\approx 13$\% of the solar value given by 
\citealt{grevesse99}). The renormalized  abundance distribution is still ambiguous:
the coronal abundance of Fe is higher than the photospheric value, but a clear FIP dependent abundance
distribution is not visible.


We also note here that the derived abundances in the SU Aur source are atypical for
young, active  stars  and T Tau stars. Their X-ray sources usually show a well-expressed 
inverse FIP effect (e.g., \citealt{telleschi06,  argiroffi04}). In contrast, SU Aur shows O and Ne 
abundances clearly lower than Fe, which is a defining signature of a solar-type
FIP effect. 
And second, as suggested earlier, the absolute Fe abundance is quite high (0.67-0.81 times
the solar photospheric value).  Such high values are usually reported for relatively
inactive stars, while magnetically active stars reveal much stronger  depletion of Fe
\citep{telleschi05, guedel04}.

\subsection{Luminosities}

The only previous report on X-ray emission from AB Aur that we are aware of is by
\citet{zinnecker94}. This allows us to study possible long-term changes in the X-ray output.
\citet{zinnecker94} give an X-ray luminosity of  $(0.3 \pm 0.09) \times 10^{30}$~erg~s$^{-1}$ in the 
energy range 0.1--2.4 keV. 
In contrast, we find  $L_{\rm X} \approx 0.4 \times 10^{30}$~erg~s$^{-1}$  in the 
energy range  0.3--10.0~keV, and  $L_{\rm X} \approx (0.5 - 0.6) \times 10^{30}$~erg~s$^{-1}$ in the 0.1--2.4 keV range,
that is, almost twice as much as \citet{zinnecker94}. We  note, however, that  \citet{zinnecker94} 
estimated $L_{\rm X}$ directly from  count rates, using a conversion factor applicable for a
temperature of 1~keV and an $N_H$ corresponding to $A_V = 0.65$~mag. Modeling the ROSAT
count rate for $N_H, kT$ and abundances found from our spectral analysis still results in only 
 $(0.3-0.4) \times 10^{30}$~erg~s$^{-1}$ in the 0.1--2.4 keV range. AB Aur was in a 
 more active state during our observation, but
 we also recall that this source is slowly variable on time scales of hours (Fig.~\ref{lc}).
Our luminosity of SU Aur is $L_{\rm X} \approx (7.4-7.8) \times 10^{30}$ erg~s$^{-1}$ in the 0.3--10~keV range.
This compares well with values reported by \citet{skinner98}, i.e.,  $(8.4 \pm 0.09) 
\times 10^{30}$ erg~s$^{-1}$  in the energy range  0.5--10.0~keV.

\subsection{Variability of spectral fit parameters}

In view of the sinusoidal variation of the X-ray count rate
in Fig.~\ref{lc}, we tested which spectral fit parameters
are mainly responsible for the modulation. We restricted
this study to the combined EPIC MOS1+2 data because the RGS
signal becomes too weak if the data are split up. We first
fitted the entire MOS data with two thermal components to
obtain a fit very similar to the one reported in Table~\ref{tab:fitres},
apart from some deviations in the element abundances
(most of which are difficult to derive from MOS, in particular
O and Ne, given the modest resolution and severe blending).
We then split the observation into  ``high state''
(first half) and ``low state'' (second half). Starting
with the above model fit, we tested whether i) an adjustment
of $N_H$ (variation due to selective absorption), ii) a renormalization
of the emission measures by a common factor (variation of
$L_{\rm X}$), or iii) some changes in all of $N_H$, $kT$, and EM
are required. A variation of $N_H$ could clearly be excluded.
This is supported by Fig.~\ref{lc} that shows that the hard
photons vary in concert with the soft photons, while they
are not significantly affected by the weak photoelectric
absorption. To quantify this, we have measured the amplitude of 
the sine function relative to the ``zero level'' in each light curve 
and found them to be the same in each energy range,
within the errors.
 No significant changes in $kT$ were required, while a simple
renormalization of the EMs by a common factor produced perfect fits.
We conclude that the variation of the X-rays are either due to an intrinsic
change in the luminosity with other plasma parameters remaining
equal, or due to an energy-independent filtering of photons (e.g.,
due to partial eclipses by the star).

\section{He-like triplets, densities, and the radiation field}\label{triplet}

We now discuss the helium-like line triplet of O\,{\sc vii} for AB Aur. The flux ratio between the 
forbidden and the intercombination lines  at 22.1~\AA\ and 21.8~\AA, respectively, is density-sensitive
roughly in the range of electron densities between $10^{10}$~cm$^{-3}$ and  $10^{12}$~cm$^{-3}$
\citep{gabriel69} for the following reason: if the electron collision rate is sufficiently high, electrons in the upper 
level of the forbidden transition, $1s2s\ ^3S_1$, do not return to the ground level, $1s^2\ ^1S_0$, instead they 
are collisionally excited to the upper levels of the intercombination transitions, $1s2p\ ^3P_{1,2}$, 
from where they decay radiatively to the ground state. They thus enhance the flux in the intercombination line
and weaken the flux in the forbidden line. However, photons in a UV radiation field may excite 
the same transition. The relevant photon energies correspond to the energy difference of the two upper
states, and this corresponds to a wavelength of 1630~\AA\ for the O\,{\sc vii} triplet. The UV radiation field
is thus important for stars with $T_{\rm eff}$ of about $10^4$~K and more. Because $T_{\rm eff}$ of AB Aur
has been quoted to be around $10^4$~K (Table~\ref{tab2}), 
we need to consider the radiation term. We follow \citet{blumenthal72} for a rough estimate.
The measured ratio $\mathcal{R} = f/i$ of the forbidden to the intercombination 
line flux can be written as
\begin{equation}\label{foveri}
\mathcal{R} = {\mathcal{R}_0 \over 1 + \phi/\phi_c + n_e/N_c} = {f\over i}
\end{equation}
where $\mathcal{R}_0$ is the limiting flux ratio at low densities ($\mathcal{R}_0 \approx 3.85$ for O\,{\sc vii}
at the maximum formation temperature), $N_c$ is the critical
density at which $\mathcal{R}$ drops to $\mathcal{R}_0/2$ ($N_c \approx 3.4\times 10^{10}$~cm$^{-3}$ for O\,{\sc vii}), 
and $\phi_c$ is the critical photoexcitation rate. The influence of the radiation field is given by 
\begin{equation}\label{phi}
{\phi\over \phi_c} = {3(1+F)c^3\over 8\pi h \nu^3} {A(1s2p\ ^3P \rightarrow 1s2s\ ^3S_1)\over A(1s2s\ ^3S_1 \rightarrow 1s^2\ ^1S_0)}u_{\nu}
\end{equation}
where the radiation field energy density is given by Planck's equation,
\begin{equation}
u_{\nu} = W {8\pi h \nu^3\over c^3} {1\over    {\rm exp}(h\nu/kT_{\rm eff}) - 1}
\end{equation}
in which we have introduced a geometric dilution factor (e.g., \citealt{mewe78}),
\begin{equation}\label{w}
W = {1\over 2}    \left(      1 - \left[         1 - \left(     {R\over d}    \right)^2       \right]^{1/2}             \right)
\end{equation} 
where $R$ is the stellar radius and $d$ is the distance of the source from the center of the
star (see also \citealt{ness01} for a more detailed presentation). At the surface, $W = 0.5$. 
Further, in Eq.~(\ref{phi}), the $A$ terms are the spontaneous transition probabilities, 
($8.12\times 10^7$~s$^{-1}$ and  $1.04\times 10^{3}$~s$^{-1}$, respectively, after
\citealt{blumenthal72}) and $F$ is an expression that we approximate by 
adopting the maximum line formation temperature for O\,{\sc vii} ($\approx 2$~MK), which yields
$F = 0.42$ for this ion \citep{blumenthal72}. We have also assumed that the radiation field exactly 
corresponds to the stellar photospheric $T_{\rm eff}$.

In a thermal plasma, the flux of the resonance line, $r$, is larger than the flux of the forbidden
line, $f$, and under conditions relevant to us, namely $T> 1.5$~MK, also the sum of $f+i$ is smaller than
$r$, $\mathcal{G} = (f+i)/r < 1$ (the ``G ratio'', see, e.g., \citealt{porquet01}). 
The considerable errors in our measurements make the extraction of individual lines and the 
separate treatment of $\mathcal{R}$ and $\mathcal{G}$ problematic. To obtain self-consistent 
$\mathcal{R}$ and $\mathcal{G}$ while fulfilling other conditions from the
spectral fit to the entire spectrum, we proceeded as follows. We adopted the optimum
parameters from the 2-$T$ fit and kept all parameters fixed, except for the electron density
and (for slight adjustments to the total O\,{\sc vii} line flux) the emission measure of the cooler component.
The thermal structure thus sets a requirement on $\mathcal{G}$ \citep{porquet01} and also fixes the
faint continuum required by other lines. We then performed a fit only to the spectral range around the 
O\,{\sc vii} triplet, in the wavelength interval 21.4--22.4~{\AA}. 

\begin{figure}
\centering
\includegraphics[angle=-90, width=0.45 \textwidth]{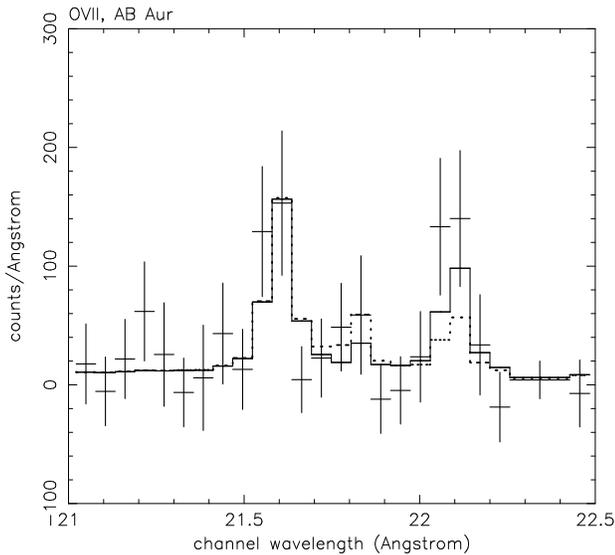}
  \caption{Fit of the O\,{\sc vii} triplet using variable electron density. The solid histogram gives the best fit
  ($n_e < 10^{10}$~cm$^{-3}$), while the dotted histogram is for the 90\% upper limit to $n_e$
  ($n_e < 1.3\times 10^{11}$~cm$^{-3}$). }
      \label{ovii}
\end{figure}

We first assume negligible influence by the radiation field, i.e., we assume that $f/i$ is controlled by the
electron density. Our fit procedure was making use of the implementation
of the He-like triplet calculations in XSPEC's vmekal code. Varying the electron density changes both the 
$i$ and $f$ flux until a best fit is obtained. We performed this procedure for various data binning
schemes, using bin widths of 45~m\AA\ or 56~m\AA. Each time, the best-fit density converged 
to values below the low-density limit for O\,{\sc vii} ($n_{\rm e} \la 10^{10}$~cm$^{-3}$, corresponding
to $\mathcal{R} = \mathcal{R}_0$). We then varied $n_{\rm e}$ to find the 68\% and 90\% upper limits.
These themselves fluctuated for different binning schemes; for the average 90\% upper limit, we find
$n_{\rm e, max, 90} \approx (1.3\pm 0.4)\times 10^{11}$~cm$^{-3}$, corresponding to the measured $f/i = 0.95$. For the 
68\% errors, we find  $n_{\rm e, max, 68} \approx (4.2\pm 1.2)\times 10^{10}$~cm$^{-3}$, corresponding to 
$f/i = 2.41$.  Fig.~\ref{ovii} shows the fit for the low-density limit (solid histogram), and the 90\% upper 
limit (dotted histogram). The strong $f$ line clearly requires a low-density environment; the 90\% limit, 
while formally acceptable, requires $i > f$, unlike the data.  The densities suggested here, 
$n_{\rm e} < 10^{11}$~cm$^{-3}$,  are very typical for stellar coronae (e.g., \citealt{ness01}).

We now consider the influence of the radiation field. UV radiation will {\it lower} 
the $f/i$ ratio and thus simulate higher densities, i.e., the electron densities 
reported above are overestimated (Eq.~\ref{foveri}). We now assume that the $f/i$ ratio is
not suppressed by high electron densities and ask how far the source must be from the star in order to
show the observed $f/i$ ratio. For $T_{\rm eff} = 10050$~K (Table~\ref{tab2}), 
we find from Eq.~(\ref{foveri})--(\ref{w}), that the 68\% upper limit ($f/i = 2.41$) 
is attained at  $d = 4.6R$, whereas the 90\% upper limit ($f/i = 0.95$)
requires  $d > 2.1R$.  For $T_{\rm eff} = 9500$~K reported by \citet{vandenancker98}, 
the 68\% upper limit corresponds to  $d = 3.6R$ and the 90\% upper limit  
to  $d = 1.7R$. For a given $\mathcal{R}$  ratio, the origin 
of most of the emission must fulfill these requirements, but some
contributions from closer to the star are not excluded. 

We have assumed here that the radiation temperature is
identical to $T_{\rm eff}$ which, due to spectral modifications,
may be inaccurate. \citet{ness01} found, for the relevant radiation at 1630~{\AA},
$T_{\rm rad}$ several hundred K below $T_{\rm eff}$ for nearby F and G-type
stars, while \citet{ness02} reported $T_{\rm rad}$ about 300~K below $T_{\rm eff}$ for the B8 star Algol
($T_{\rm eff}$ = 13000~K). But even if we adopt $T_{\rm rad} = 9000$~K for AB Aur,
we still require $d >2.8R$ and $d>1.3R$ for the 68\% and the 90\% limit
of the $f/i$ ratio.

In conclusion, we find that the electron density in the source must not exceed a few 
times $10^{10}$~cm$^{-3}$ (about $10^{11}$~cm$^{-3}$ for 90\% upper limit) and the 
majority of the source plasma must not be closer to the stellar center 
than $(1.3-2.1)R$, depending on the exact temperature of the  radiation field.

\section{Discussion}\label{discussion}

In Table~\ref{tab2} we have listed the basic properties of AB Aur and 
we have compared them with those of the CTTS SU Aur and HD 163296, an 
Herbig star that has been reported to reveal a soft spectrum \citep{swartz05}.

The similarity between AB Aur and HD~163296 is 
striking. The stars not only have
very similar fundamental properties, but their X-ray properties are also very similar. In contrast, SU Aur, 
a relatively massive but markedly later-type classical T Tau star, shows X-ray properties 
very different from the Herbig stars, although these properties are typical among T Tau stars \citep{guedel06}.

We now discuss various models for X-ray production proposed in the
previous literature and compare predictions made by these models with
our spectroscopic results. We largely follow the presentation by
\citet{skinner96}.

\subsection{Atmospheric and wind structure}\label{atm}

Numerous optical and ultraviolet spectral observations of AB Aur have converged
to a wind+chromosphere model in which an expanding wind overlies 
an extended, hot chromospheric layer. The latter has a height of
about 1.5$R$, with a temperature peak of $1.7\times 10^{4}$~K
\citep{boehm93}. \citet{praderie86} and \citet{catala86} discovered
periodic modulation in the Mg\,{\sc ii} and O\,{\sc ii} lines, although the periods
disagreed, the period of Mg\,{\sc ii} being 45~hr and the period of O\,{\sc ii}
being near 32~hr. The latter was identified with the stellar rotation
period, implying formation of O\,{\sc ii} close to the star, while Mg\,{\sc ii}
forms further away from the star (in its wind at several $R$) and may 
therefore be modulated by the rotation period of the envelope {\it at that distance}.
The modulation was suggested to be due to a non-axisymmetric  wind
in which fast and slow streams alternate. Magnetic fields would then provide
a possible explanation to this wind structure. High-temperature lines 
of N\,{\sc v} and O\,{\sc vi} were detected by FUSE, in AB Aur \citep{bouret97, roberge01} 
and the similar HD~163296 \citep{deleuil05}, and these were interpreted 
as originating from shocks formed when the fast and slow winds collide
\citep{bouret97}. The same model could also produce X-rays in a layer close to 
the star (0.05$R$ above the surface, \citealt{bouret97}).
However, the $f/i$ ratio that we measured in the X-ray spectrum suggests
that the X-rays are formed at distances  d $>$ 1.3 $R$ from
the center of the star (for $T_{\rm rad} > 9000$~K).
Alternatively, N\,{\sc v} and O\,{\sc vi} could also be produced in a wind shock
along with X-rays in the model by \citet{babel97} described  below.

\citet{catala99} extended periodicity studies to photospheric lines
and found that the amplitudes of the red emission components are modulated  with a period of 43~hr,
whereas the {\it velocity} of the blue absorption components is subject to a 34~hr period.
Further, the HeI~D3 line shows red and blue components that are both
modulated with a period around 45~hr (red in amplitude, blue in velocity). 
These authors interpreted the blue HeI~D3 components as originating from an equatorial wind,
whereas the photospheric blue absorption would come from high-latitude
photospheric regions with radial flows, indicating a shorter period at 
high latitudes than near the equator. Finally, all redshifted  components 
would be due to polar infall. Their 43--45~hr period remains unexplained, but 
the infall may be magnetically linked to outflows at lower latitudes. 
Alternatively, the 45~hr HeI (blueshifted) outflow signatures may originate 
from a magnetic disk wind, indicating an anchor point of the magnetic fields 
at 1.6$R$ \citep{catala99}.

The above interpretation is subject to one caveat, namely the assumed high
inclination angle (70$^{\circ}$ in \citealt{catala99}). Adopting 
a radius of $2.3~R_{\odot}$ (Table~\ref{tab2}), $v\sin i = 80$~km~s$^{-1}$ 
(\citealt{boehm93}, Table~\ref{tab2}), and the extreme value of $i = 70^{\circ}$
(see also references in 
\citealt{corder05}), we find $P = 33$~hr, in perfect agreement with 
the 32-34~hr period reported from O\,{\sc ii} and blue photospheric absorption
(for $i = 90^{\circ}$, the maximum rotation period is 35~hr). 
However, recent observations have strongly revised the {\it disk} inclination
angle and now suggest $i \approx 21.5^{\circ}$ (\citealt{corder05} and further 
references therein). This suggests  $P = 12.9$~hr. The 32-34~hr {\it stellar}
rotation period could only be maintained if the disk and stellar axis
were grossly mis-aligned.

It is interesting that our X-ray light curve period of $42.5\pm 4$~hr 
perfectly agrees with the period in Mg\,{\sc ii} and He~I, i.e., components 
formed in the chromosphere at the wind base and in the wind itself, but
is clearly not compatible with the rotation period based on $v\sin i$ 
measurements with $i \approx 21^{\circ}$, or the photospheric blue absorption components.  
We thus tentatively conclude that the X-ray production may be related,
in some ways, to the wind.

\subsection{Wind shocks}

In hot stars, shocks are driven by instabilities in line-driven winds (see, for
example, \citealt{feldmeier97a} for a review). The electron temperature of AB Aur's
X-ray source is similar to those measured in O stars \citep{feldmeier97a}. The important
parameter for the shock instability is the Eddington parameter
\begin{equation}
\Gamma = {\sigma_e\over 2.5\times 10^4}  { L_*\over M} 
\end{equation}
\citep{castor75} where we set the electron mass scattering coefficient
$\sigma_e = 0.4$~cm$^2$~g$^{-1}$, and $M$ is the stellar mass. We thus obtain
$\Gamma = 5.6\times 10^{-4} \ll 1$. Under these circumstances,
the wind cannot be radiation-driven, which makes the instability-shock
hypothesis unlikely.

On the other hand, \citet{zinnecker94} have suggested that the wind slams into
dense molecular material  in the ambient medium. Using an average wind velocity of
\citet{catala87}, $v_w = 225$~km~s$^{-1}$, and the wind mass loss rate reported by
\citet{skinner93}, $\dot{M}_w = 1.1\times 10^{-8}~M_{\odot}$~yr$^{-1}$ from radio
observations, we derive a ``kinetic luminosity'' $\dot{M}_wv_w^2/2 \approx
1.8\times 10^{32}$~erg~s$^{-1}$. That means that only about 0.3\% of the available wind
energy needs to be dissipated in shocks to produce the X-ray luminosity.

However, the observed systematic variability of the X-ray source on time scales of
several hours makes models based on very-large-scale shocks as well as on many distributed
shocks in the wind of a single star unlikely. This echoes the conclusions by \citet{skinner05}
who summarize observations of X-ray variability in O stars. Although \citet{feldmeier97b}
proposed variability owing to colliding shells in the O-star wind,  the time scales for 
such collisions would be shorter (of order 500~s).  

\subsection{Magnetically confined winds}\label{magn_wind}

Although strong magnetic fields are not expected on Herbig stars given their predominantly
radiative interior, large-scale fossil magnetic fields from the stellar formation process could still
be trapped in the star. There have been a number of investigations that studied
the consequences of ionized winds trapped in large-scale stellar magnetic fields.

\citet{havnes84} presented a model of a global stellar magnetosphere which is
fed by a wind from the stellar surface. They considered a magnetic field with a 
dipole configuration, where plasma is confined by closed magnetic field lines. 
Close to the stellar surface, where the gravitational forces
exceed the centrifugal forces, the density is low, while it increase further
out. The temperature also increases, possibly to coronal
temperatures, at 5-10 stellar radii. The outward transport of plasma takes place 
by events where the magnetic lines are broken due to excessive density.
The energy input in the corona from rotation is given by $\dot{E}= 1/2~\dot{M}_w  
\Omega^2 R^2 (L_2^2-L_1^2)$, where $\Omega$ is the angular velocity. 
This can be rewritten as $\dot{E} \approx 1.7 
\times 10^{17} \dot{M}_w R^2(L_2^2-L_1^2)~P_{\rm d}^{-2}$~erg~s$^{-1}$ (where $\dot{M}_w$ 
is in $M_{\odot}$~yr$^{-1}$, $R$ is in cm and the rotation period of the star $P_{\rm d}$ 
is in days). The parameters $L_1$ and $L_2$ are the distances to the inner and outer edges of the 
X-ray emitting region in units of the stellar radius. Both $L_1$ and $L_2$ are a 
function of the mass loss, the magnetic field and the rotation period, and are 
therefore difficult to estimate. \citet{havnes84} estimated these values to
be $L_1=15$ and $L_2=20$, for a hotter and more massive star. Even if we cannot
constrain these two values, we expect that ($L_2^2-L_1^2$) will be
larger than 1, so that we obtain $\dot{E} \ge 4 \times 10^{31}$~erg~s$^{-1}$,
i.e., enough to produce the observed $L_{\rm X}$.


In another model, \citet{usov92} considered the wind zone (open field lines) outside the 
corotating magnetosphere (dead-zone). 
According to their model a current sheet is formed in the equatorial plane outside the dead 
zone that separates regions with opposite directions of the magnetic field. The authors 
estimated the temperature and the energy released in the current sheet, assuming bremsstrahlung 
as a cooling agent.
Assuming a surface magnetic field strength of 100~G (see recent measurements on Herbig stars by
\citealt{hubrig04}), the above $\dot{M}_w$, an average wind velocity of $v_w = 225$~km~s$^{-1}$
\citep{catala87, skinner93},  and a radius of $2-4~R_{\odot}$, we find for the two parameters $\xi$ and
$\eta$ in their Eqs. (5) and (23),  $\xi \gg 1$ and $\eta \ge 1$, respectively, and therefore, from
Eq. (25) in \citet{usov92}, an X-ray luminosity of $1.1\times 10^{30}$~erg~s$^{-1}$. Their Eq. (20) predicts
an electron temperature of $T \approx 6\times 10^7 (v_w/10^8~{\rm cm~s^{-1}})^2 \approx 3\times 10^6$~K.
These parameters are again in good agreement with our measurements.


Finally, \citet{babel97} considered the wind shock inside the magnetosphere that
emerges when the magnetically guided winds from the northern and the southern
hemispheres collide in the equatorial plane. They predict a shock temperature of
$T_s = 1.13\times 10^5 (v_w/10^7~{\rm cm~s^{-1}})^2 \approx 0.5-1$~MK for  
$v_w = 200-300$~km~s$^{-1}$ (extreme values reported by \citealt{catala87}).
Magnetically guided winds develop shocked equatorial ``disks''  only if the 
magnetic fields are sufficiently strong for confinement. For this, the (equatorial) wind confinement
parameter, $\eta = B^2 R^2/(\dot{M}v_W)$ must be at least unity (\citealt{uddoula02}; note that $B$ is the
surface magnetic field). With our stellar parameters (Table~\ref{tab2}) and $B = 100$~G, we find 
$\eta \approx 20$, and hence wind shocks can develop. The temperatures provided by this model are 
nevertheless too low with respect to the observations.

In summary, wind-fed magnetospheres may be promising to produce the observed X-ray emission
although details of the magnetic field arrangement and the wind-field interactions would need
further elaboration for the specific case of AB Aur.

\subsection{Accretion-induced X-rays}

Accretion has recently gained some attention as a possible contributor to
X-ray emission in classical T Tau stars. Accretion shocks at the base of
magnetic funnel flows may reach high temperatures, and
high densities (of order $10^{13}$~cm$^{-3}$, \citealt{ulrich76}).
In standard accretion shock models, the shock heats up to a few times $10^6$~K,
with the ensuing X-rays heating the underlying photosphere to produce an UV excess \citep{calvet98}.
\citet{lamzin99} concluded that a typically small fraction of the X-rays escape from the shock
that can be seen in soft X-rays. However, it will be important to estimate whether
the shock is above the photosphere at all and therefore visible, and this is not normally
the case for T Tau stars with average accretion characteristics \citep{calvet98}.
Little variability should  be seen in $L_{\rm X}$ or the electron temperature in shocks  
\citep{lamzin99}. \citet{kastner02} have proposed accretion-induced X-ray production 
for the unusually soft X-ray emission and the high densities measured in the spectrum of the 
classical  T Tau star TW Hya. In analogy, \citet{swartz05}
suggested the same scenario for the soft spectrum of the Herbig star
HD~163296 (Table~\ref{tab2}), but high-resolution X-ray spectroscopy was not available for this
star.

A further argument in favor of an accretion model was brought up by \citet{stelzer04}
who argued that TW Hya's anomalously high abundances of Ne and N support 
an accretion scenario; depleted metals would condense onto grains 
in the disk, leaving gas enriched in certain other elements, and 
this gas would eventually accrete onto the star. 
However, refering to Ne/Fe and N/Fe abundance ratios, there are also
evolved stars and non-accreting pre-main sequence stars that reveal high values
for these ratios (see \citealt{guedel04} for a review). As for the Ne/O abundance ratio, \citet{drake05}
indeed found it to be unusually high in TW Hya compared to other stars.
However, such a high Ne/O abundance ratio has so far been measured only
in TW Hya. In other accreting and non-accreting stars, this ratio is found to be
half as high as in TW Hya \citep[see for example][]{argiroffi05,robrade06,telleschi06}, 
and consistent with values of a large sample of late-type stars \citep{drake05,telleschi05}.
\citet{drake05} suggested that the anomalous Ne/O abundance in TW Hya is determined by
the higher degree of metal depletion in this older star. This ratio can therefore
not in general be used as an accretion signature in young and less evolved TTS.

What do we know about accretion in Herbig stars?
Evidence for accretion from a disk
is rather indirect. The temperature of the photosphere that is heated by infalling material
is very similar to the undisturbed photosphere \citep{muzerolle04}. There has been some evidence
for mass inflow in Herbig stars from redshifted absorption components
in optical lines \citep{sorelli96, natta00, catala99}, but the interpretation is model-dependent, with $\dot{M}$ possibly exceeding
$10^{-7}~M_{\odot}~{\rm yr}^{-1}$ \citep{sorelli96}. \citet{blondel93} suggested that hydrogen 
Ly$\alpha$ lines in Herbig stars are due to infalling gas, although no signatures were
found for AB Aur. \citet{grady99} discussed evidence for
infalling gas in the case of AB Aur. \citet{muzerolle04} interpreted Balmer
and sodium profiles based on magnetospheric accretion models to conclude that $\dot{M} \approx
10^{-8}~M_{\odot}~{\rm yr}^{-1}$ for the Herbig star UX Ori, and for a larger sample of stars
that excess fluxes in the Balmer discontinuity imply $\dot{M} \la 10^{-7}~M_{\odot}~{\rm yr}^{-1}$.
However, the  Balmer discontinuity excess has been measured to be $0\pm 0.1$~mag in 
AB Aur \citep{garrison78}, at best implying $\dot{M} \approx 10^{-8}~M_{\odot}~{\rm yr}^{-1}$
(after \citealt{muzerolle04}). This is supported by measurements by \citet{boehm93} who find
$\dot{M} \la 7.5\times 10^{-8}~M_{\odot}~{\rm yr}^{-1}$. \citet{catala99} find explicit evidence for
near-polar downflows in AB Aur with velocities of about $300$~km~s$^{-1}$, leading to estimates for
$\dot{M}$ of a few times $10^{-9}~M_{\odot}~{\rm yr}^{-1}$ (see also \citealt{bouret00}).

The accretion model is specifically favored for AB Aur by the measurements of the red component of the He~I D3
line by \citet{catala99}. In fact, they found a periodicity in the redshifted line amplitude
of 42--45 hr, i.e. very close to the period that we measure in the X-ray light curve.

We first need to check whether an accretion rate of order $\dot{M} \approx 10^{-8\pm 1}~M_{\odot}~{\rm yr}^{-1}$
suffices to explain the observed X-ray output if the latter is indeed generated by accretion shocks.
The accretion luminosity is, ignoring  viscous dissipation in the disk,
$L_{\rm acc} = GM\dot{M}/(2R) \approx 6\times 10^{40}~\tilde{M}\dot{M}/(M_{\odot}~{\rm yr^{-1}})/\tilde{R}$~erg~s$^{-1}$
or $L_{\rm acc, 30} \approx 600\tilde{M} \dot{M}_{-8}/\tilde{R}$ where 
$L_{\rm acc, 30} = L_{\rm acc}/(10^{30}~{\rm erg~s^{-1}})$, $\tilde{M} = M/M_{\odot}$, $\tilde{R} = R/R_{\odot}$,
and $\dot{M}_{-8}$ is  $\dot{M}$ in units of $10^{-8}~M_{\odot}~{\rm yr^{-1}}$
(similar  values were reported for  HD~163296 given the very similar parameters for this
star and AB  Aur - see Table~\ref{tab2}, \citealt{swartz05}). We conclude that there is sufficient accretion 
energy available to produce the X-rays. 

We now estimate what an accretion model would predict for X-ray production on
AB Aur.  The expected shock temperature is, from the strong-shock conditions, $T = 3v^2\mu m_p/16k$
where the upstream flow velocity is approximately  equal to the free-fall velocity,
$v_{\rm ff} = (2GM/R)^{1/2}$ (ignoring centrifugal forces if mass is guided along rotating
magnetic fields), $m_p$ is the proton mass, and $\mu \approx 0.62$ for a fully ionized gas.
We thus find
\begin{equation}
T \approx 5.4\times 10^6 {\tilde{M}\over \tilde{R}}~{\rm [K]}
\end{equation}
and with the parameters in Table~\ref{tab2}, $T \approx 6$~MK, in good agreement
with our measurements.

To estimate the shock density, we use the strong-shock condition $n_2 = 4n_1$ where 
$n_1$  and $n_2$ are the pre-shock and post-shock densities, respectively. We first estimate 
$n_1$ from the total mass accretion rate and the estimated accreting area on the surface: 
$\dot{M} \approx 4\pi R^2 f v_{\rm ff} n_e m_p$ 
where $f$ is the surface filling factor of the accretion flows, or 
$\dot{M}_{-8}\approx 1 \times 10^{-11} \tilde{R}^{3/2}\tilde{M}^{1/2} f n_1$. We thus find\footnote{In 
  principle, $n_2$ could also be derived from the observed $L_{\rm X}$ and the shock volume, which
  derives from $f$ and the shock height, the latter being dependent on the inflow velocity and the
  cooling time, see \citet{calvet98}. However, such shock models show that the X-rays are
  attenuated by 1--2 orders of magnitude due to the infalling material, see \citet{lamzin99}.
  Assuming that all $L_{\rm X}$ escapes would, when scaled with the observed luminosity, lead
  to an underestimated $n_2$ and an overestimated shock height.} 
\begin{equation}
n_2 \approx {4 \times 10^{11}\over \tilde{R}^{3/2}\tilde{M}^{1/2}} {\dot{M}_{-8} \over f}~{\rm [cm^{-3}]}.
\end{equation}
For the sake of argument, we adopt $\dot{M}_{-8} = 1$,
which seems to be a reasonable value based on the findings summarized above, and
which is similar to accretion rates of less massive T Tau stars. For typical filling factors 
as discussed by \citet{calvet98}, i.e., $ f = 0.1-10$\%, we find 
$n_2 \approx 10^{12}-10^{14}$~cm$^{-3}$.

Although these are densities similar to those measured on the 
T Tau star TW Hya, the O\,{\sc vii} triplet we see in AB Aur requires densities about 100 times 
smaller. This could only be achieved by lowering $\dot{M}$ to about $10^{-10}~M_{\odot}$~yr$^{-1}$ 
or by increasing the accretion area to essentially the entire stellar surface. The former possibility
is not supported by (at least tentative) measurements of limits to $\dot{M}$, as summarized above.
Accretion onto the entire stellar surface is unreasonable given that the star accretes from a 
disk, and a wind is present (e.g., \citealt{praderie86}).

We have not yet considered the radiation field of the A star. As shown in Sect.~\ref{triplet},
a minimum distance of $1.3-2.1R$  from the center of the star
 is required for the X-ray source to be compatible with the observed
O\,{\sc vii} $f/i$ flux ratio. One way out is that the shocks are sufficiently shielded from UV radiation.
It is not clear, then, how X-rays can escape without any absorption in addition to the
circumstellar photoelectric absorption that agrees well with the optical extinction from 
circumstellar dust. This absorption is in fact extremely low in AB Aur, compared to other 
young stars in the Taurus-Auriga molecular cloud \citep{guedel06}.

In summary, then, certain properties of X-ray production in accretion shocks may 
well be explained by simple shock models, but there are serious problems with
this explanation, in particular related to i) the radiation field, ii) the low 
densities measured in the O\,{\sc vii} triplet and iii) the lack of any excess absorption. There is
also little support with regard to selective condensation of metals onto grains in
the accretion disk. The abundance of Fe, thought to be among the elements that condense easily
onto grains \citep{stelzer04}, is {\it higher} in the X-ray source than in the  photosphere 
\citep{acke04}.

\subsection{Coronal X-rays}\label{coronae}

Coronae provide the standard explanation for X-rays from stars of spectral type F and
later, because these stars maintain an outer convection zone that drives a dynamo. AB Aur is, however,
a late-B or early-A type star. 

Models of Herbig stars have indicated that a transient shell in which deuterium is burned may develop
in Herbig stars, although this occurs preferentially in the later-type Herbig  Ae stars  \citep{palla93}.
The calculations of \citet{siess00} of pre-main-sequence evolutionary tracks also predict the presence 
of a thin convective layer in young AB stars. Using their evolutionary model, we obtain for AB 
Aur a thin convection zone of 0.2\% of the stellar radius.

The situation resembles that of mid-to-late A type stars and early-F type stars on the main 
sequence.  Although some of these stars are X-ray sources, they are clearly subluminous 
compared to later-type stars (if normalized with $L_{*}$) despite their often rapid 
rotation, and the X-ray spectra are soft \citep{panzera99}. It appears that the thin 
convection zones of these stars are unable to maintain vigorous magnetic dynamos, resulting 
in soft, solar-like coronae. A similar situation may apply to Herbig stars like AB Aur.

The average temperature of AB Aur is similar to $T_{\rm av}$ of moderately active 
main-sequence solar analogs (G2-5~V), such as $\pi^1$ UMa, $\chi^1$ Ori, and $\kappa^1$ Cet
\citep{telleschi05}. Assuming that AB Aur reveals a similarly structured corona, we
expect that $L_{\rm X}$ scales with the surface area. Adopting $R = 2.3R_{\odot}$ for AB Aur,
 its $L_{\rm X}$ would be $\approx (5-6)\times 10^{29}$~erg~s$^{-1}$, in agreement
 with the observations.
We also note that the densities derived from the O\,{\sc vii} lines are in good agreement
with  measurements reported for numerous coronal sources \citep{ness04}.
Finally, the modulation of the X-ray light curve observed in Fig.~\ref{lc} 
could be due to rotation and is thus also consistent with the coronal hypothesis.

\citet{tout95} have proposed a non-solar dynamo that could operate in rapidly rotating
A-type stars based on rotational shear energy. The model predicts, for the time development
of the X-ray emission from the associated corona,
\begin{equation}
L_{\rm X}(t) = L_{{\rm X}, 0}\left( 1 + {t\over t_0}\right)^{-3}
\end{equation}
where $L_{X, 0}$ is a quantity that depends on stellar mass, radius, the amount of
differential rotation in the stellar interior, the breakup velocity, the coronal heating
efficiency, and the efficiency of magnetic field generation. Following the argumentation
and the choice of constants given in \citet{skinner04} and \citet{tout95}, we find 
$L_{{\rm X}, 0} \approx 1.6\times 10^{31}$~erg~s$^{-1}$. For AB  Aur, Eq. (3.15) in \citet{tout95} 
gives, using the same default parameters, $t_0 = 6.3\times 10^5 \tilde{M}^{-1/2} 
\tilde{R}^{3/2}~{\rm yr} \approx  1.5\times 10^6$~yr.  Given the star's age of 4~Myr 
\citep{dewarf03}, we expect $L_{\rm X} \approx (3-4)\times 10^{29}$~erg~s$^{-1}$. Using our updated
parameters for AB Aur, we thus confirm the conclusion by \citet{skinner04} that this model
provides very good agreement with the observations of AB Aur, although this is not true
for most other Herbig stars.  As the star evolves, the X-ray generation  would rapidly 
decay further.

We found a coronal Fe abundance that is at least equal to the photospheric abundance. 
This is again consistent with a coronal model. In more evolved  magnetically active 
stars, the Fe abundance generally increases toward lower activity and becomes comparable
with the photospheric abundance in inactive stars \citep{guedel04}. 

We note, however, that centrifugal forces exceed gravitational forces at
a distance of 1.68 $R$ from the center of the star 
according to parameters in Table~\ref{tab2} and assuming an
(unconfirmed) rotation period of 12.9 days. For equatorial magnetic fields, then, 
the coronal radius must be less than 1.68 $R$, otherwise the loops will be
unstable \citep{collier88}. This condition is only in marginal agreement with the lower
limit to the source size from the $f/i$ ratio (Sect.~\ref{triplet}) unless
$P_{\rm rot}$ is larger.

\subsection{X-rays from a companion?}\label{binaries}

A majority of stars form in multiple systems, and this is particularly true for 
Herbig stars \citep{feigelson03}. A companion may therefore also be responsible for the observed 
X-ray emission in AB  Aur. Such a companion would most likely  be a T Tau star. 
Recently, \citet{stelzer06a} have studied a sample of 17 Herbig Ae/Be stars
with {\it Chandra}, concluding that at least in 7 stars the X-rays could
originate from an unresolved companion. Only 6 stars are found to be X-ray emitters
with no visual or spectroscopic detection. Furthermore, the X-ray properties
in this stellar sample are very similar to X-ray properties of CTTS.   

\citet{behar04} postulated that the X-ray emission detected
from the late-type B star $\mu$ Lep in fact originates from an unknown pre-main-sequence 
companion, given the high $f/i$ ratio measured in its spectrum. For AB Aur, the $f/i$ ratio
requires the bulk  X-ray emission to 
originate at $r > 1.38 R$. This would be fulfilled if the X-ray emission originated
from a companion.

Stringent constraints have
been discussed in the literature, interpreted and summarized by \citet{pietu05} (see their Sect.~5.1, and
references therein):
Any co-eval companion within 120-1500~AU (0.86--12.5$\arcsec$) must have a mass $< 0.02M_{\odot}$.
The mass upper limit is 0.25$M_{\odot}$ down to a separation of 0.4$\arcsec$. In the range of 0.07--10$\arcsec$,
the upper limit to a companion mass is, from speckle interferometry, 0.05--0.3$M_{\odot}$. 
Recently, \citet{baines06} reported evidence for binarity of several
Herbig stars, including AB Aur. However, they estimated a separation
of 0.5-3.0$\arcsec$ for AB Aur.

The point-spread function of {\it XMM-Newton} does not allow us to distinguish between
our target source and potential companions within a few arcsec. We have therefore analyzed 
an exposure of the region around  SU Aur + AB Aur obtained from the archive
of the {\it Chandra X-Ray Observatory}, revealing much better positional information
(obs ID $ = 3755$). Standard data reduction methods  and up-to-date
aspect corrections were applied.  The observation used the ACIS-S detector, with the high-energy
grating inserted. It was centered on SU Aur, with AB Aur being located close to the chip edge.
Using the wavdetect (wavelet  detection) routine in the {\it Chandra}  CIAO software, we measured 
the centroid positions of the images of both stars, to find 

\medskip

\noindent \begin{tabular}{lllll}
      AB  Aur &    RA(J2000.0)       & = &  4$^{\rm h}$ 55$^{\rm m}$ 45.846$^{\rm s}$ &  $\pm 0.17\arcsec$\\
              &    $\delta$(J2000.0) & = & 30$^{\circ}$ 33$\arcmin$ 04.138$\arcsec  $ &  $\pm 0.17\arcsec$\\
\end{tabular}

\medskip

\noindent \begin{tabular}{lllll}
      SU Aur  &    RA(J2000.0)       & = &  4$^{\rm h}$ 55$^{\rm m}$ 59.389$^{\rm s}$ & $\pm 0.11\arcsec$\\  
              & $\delta$(J2000.0)    & = & 30$^{\circ}$ 34$\arcmin$ 01.297$\arcsec $  & $\pm 0.11\arcsec$\\
\end{tabular}
\medskip

\noindent The nearest 2MASS objects are located at 
\medskip

\noindent \begin{tabular}{lllll}
      AB Aur  &    RA(J2000.0)       & = & 4$^{\rm h}$ 55$^{\rm m}$ 45.826$^{\rm s}$   & $\pm 0.07\arcsec$\\  
              &    $\delta$(J2000.0) & = & 30$^{\circ}$ 33$\arcmin$  04.37$\arcsec$    & $\pm 0.07\arcsec$ \\ 
\end{tabular}

\medskip

\noindent \begin{tabular}{lllll}
      SU Aur  &    RA(J2000.0)       & = & 4$^{\rm h}$ 55$^{\rm m}$ 59.381$^{\rm s}$     & $\pm 0.06\arcsec$\\  
              & $\delta$(J2000.0)    & = & 30$^{\circ}$ 34$\arcmin$ 01.56$\arcsec$       & $\pm 0.06\arcsec$\\
\end{tabular}
\medskip

\noindent The position offsets of the X-ray sources are thus 0.258$\arcsec$ and 0.103$\arcsec$ 
in RA for AB Aur and SU Aur, respectively, and $-0.232\arcsec$ and $-0.262\arcsec$ in 
declination.  The similar offsets for both stars suggest a systematic offset of the 
pointing of order 0.28$\arcsec$, well within the errors of the {\it Chandra} attitude 
solution.\footnote{The  90\% source location error circle in {\it Chandra} has a 
radius of about 0.5$\arcsec$, see {\it Chandra} Proposers' Observatory Guide v.8}
Correcting the AB Aur position by the offset of the better determined SU Aur position, 
the deviation from 2MASS of AB Aur is only 0.16$\arcsec$ in RA and 0.03$\arcsec$ in 
declination, which is within the errors of the measurement.
We conclude that we have identified AB Aur in X-rays well within 0.5$\arcsec$ (at 
the 3$\sigma$ level) of the 2MASS position. It is therefore improbable that the 
source of the X-rays is the companion detected by \citet{baines06}, which is thought to have a 
separation with AB Aur of $> 0.5\arcsec$. 

Further arguments favor intrinsic X-ray emission from the 
Herbig star. In particular, the observed low average
temperature is rather uncommon to lower-mass
T Tau stars; the latter rather show hot components
with characteristic temperatures up to 20-30 MK
as, for example, SU Aur but also other CTTS in the
XEST survey \citep{telleschi06}. Very-low mass stars and 
brown dwarfs do reveal softer spectra with dominant 
temperatures below 10 MK but their total X-ray luminosities are
much below $L_{\rm X}$ measured here for AB Aur (see \citealt{grosso06}
for the brown dwarf sample from the XEST survey).

We also note that the $N_H$ value determined in the XEST survey  
($N_H \approx 5\times 10^{20}$~cm$^{-1}$, \citealt{guedel06}) is in perfect agreement
with the visual extinction measured for AB Aur by \citet{roberge01} (0.25~mag), if we apply 
a standard interstellar conversion law, $N_H \approx 2\times 10^{21}~A_V$~mag (\citealt{vuong03} and
references therein).

The most substantial argument against the companion hypothesis is the close coincidence
between the X-ray period and the period observed in the lines of the wind of the Herbig star.

Finally, AB Aur and HD 163296 are very similar in both their intrinsic properties and
their X-ray properties. 
If the X-rays would indeed originate from nearby T Tauri companions, then the companion of AB Aur would
happen to be very similar to the companion of HD 163296. 

Taken together, these arguments suggest that the X-rays are not originating from a companion but from
AB Aur itself. This is different from flaring, hard sources among Herbig stars for which T Tau 
companions have recently been identified (see Sect.~\ref{introduction}). We therefore suggest 
that the unusually soft emission is indeed a distinguishing property of genuine Herbig star X-ray emission.

\subsection{Disk related models}

The X-ray luminosities of A and B stars decrease as they  approach and reach the main sequence
\citep{stelzer06a}; 
at the same time the outflow activity is believed to cease and the dense surrounding 
gas dissipates.
That suggests that the X-rays of AB Aur could be related to the presence of the 
circumstellar disk. Two different models that link the X-ray emission with the 
presence of the disk have been discussed in the literature: a disk corona and 
reconnection of magnetic fields that link the star to the disk.
We describe these models, although they do not currently make
predictions that we can test against our data.

The presence of a disk corona was discussed by \citet{zinnecker94}, but 
very little is known about its generation.
The ionization due to the decay of radioactive nuclides could 
increase the conductivity of the disk to a level high enough to generate a magnetic field. 
The differential rotation in the disk could then generate a disk corona. 

X-ray emission generated by reconnection of magnetic fields linking the star
to the disk has been discussed as a model for the X-ray generation mechanism 
of low-mass protostars \citep{montmerle00}. If the rotation period of the star is not the
same as the rotation period of the disk, a magnetic loop connecting the disk with the star will twist
and inflate until it comes into contact with itself and reconnects.
A similar model could apply to Herbig stars, although we would expect 
higher, flare-like temperatures due to the reconnection process.

\bigskip

\section{Conclusions}\label{conclusions}

We have presented the first high-resolution X-ray spectrum of an Herbig Ae/Be
star, namely AB Aur. 
The use of high-resolution spectroscopy has allowed us to obtain important spectral information
that cannot be addressed with  EPIC spectra alone. The O\,{\sc vii} triplet
constrains the electron density and is therefore important for the discussion of the different models. Further,
we have been able to reliably determine the abundances of the high-FIP elements O and N. Finally, the
O\,{\sc viii}, O\,{\sc vii} and N\,{\sc vii} lines permitted us to constrain the cool plasma.

We found the X-ray spectrum of AB Aur to be  rather soft, with spectral-fit
results that are consistent with a mean coronal temperature of about 5 MK,
i.e., much less than the usual temperatures of coronae of low-mass pre-main-sequence
stars that usually exceed 10 MK.
We found an X-ray luminosity of about $4 \times 10^{29}$ erg/s in the
0.3--10 keV range. 
We derived the abundances and found them not to follow a First
Ionization Potential (FIP) distribution, nor an inverse FIP distribution.
We normalized the coronal abundances to the new photospheric abundances of AB Aur
found by \citet{acke04}, who measured a very low photospheric Fe abundance. The Fe coronal
abundance then is at least as large as the photospheric value.

The density-sensitive O\,{\sc vii} triplet has been studied in detail. Although its
S/N is moderate, we found that the line flux ratios indicate densities $n_{\rm e} \la 10^{11}$ cm$^{-3}$,
with the best-fit value being at the low-density limit ($n_{\rm e} \approx 10^{10}$~cm$^{-3}$),
similar to what is commonly found in stellar coronae \citep{ness04}.

We have discussed several X-ray generation mechanisms,  and provided supporting
evidence or pointed at problematic features for each of them. First, the probability
that the X-ray emission is originating from a companion TTS is small. 
The X-ray source is identified within 0.5$\arcsec$ of the 2MASS source corresponding to AB Aur in the 
{\it Chandra} image, so that the companion would have a mass  $< 0.3M_{\odot}$ according to the
constraints given by \citet{pietu05}. Such a low-mass star would rarely
produce an X-ray luminosity as high as observed for AB Aur. Furthermore,
the close coincidence of the period that we have measured in X-rays with
the period measured in lines formed in the wind of AB Aur itself, makes
the companion hypothesis unprobable.

Accretion-induced X-ray emission has been widely discussed
in the recent literature. With the observed electron density, 
this would be possible only if $\dot{M} \approx 10^{-10}~M_{\odot}~{\rm yr}^{-1}$,
i.e. lower than the value suggested in the literature 
($\dot{M} \approx 10^{-8\pm 1}~M_{\odot}~{\rm yr}^{-1}$),
and a filling factor $\ga 10$\%. The major problem with this 
scenario is that the UV radiation field (of the stellar photosphere 
and the shock itself) would suppress the forbidden line of the O\,{\sc vii} triplet.


The hypothesis that X-ray emission is generated by shocks in a line-driven wind,
similarly to the mechanism that is believed to produce X-rays in O stars, is
ruled out by the observed variability and the inability of the radiation
field to drive the wind.

Magnetic fields have recently been detected on several Herbig stars \citep{donati97, hubrig04, wade05}.
Two further possible mechanisms are fundamentally dependent on the existence of a magnetic field: 
coronal emission and magnetically confined winds.

Coronal emission of the type seen in the Sun  requires the presence of a dynamo. 
The recent calculations by \citet{siess00} predict a thin convective layer for Herbig stars, which
in the case of AB Aur  amounts to 0.2\% of the radius.
The question then is whether this thin
convection layer would be sufficient to generate the dynamo that results in a
corona with the observed X-ray properties. 
The corona should be quite extended in order to allow
the $f/i$ flux ratio in the O\,{\sc vii} triplet to be larger than unity.  This is plausible because 
the surface gravity of AB Aur  is about half that of the Sun.

Magnetically confined winds are a promising alternative the more so that winds have explicitly been
measured \citep{catala99}. 
There is interesting evidence supporting a model of this kind. The period of the modulation that we 
measured in the X-ray light curve (42.2 hr, Sect.~\ref{sec:lc}) is very close to a modulation period of 
Mg\,{\sc ii} lines thought to form in the wind \citep{praderie86}. Whatever the production mechanism
of the X-rays, it is therefore suggestive to assume that they are closely related to the
wind or are produced by it. The advantage of such a model is that X-rays are produced at some
distance from the stellar surface, which alleviates the problem with the suppression of the O\,{\sc vii} 
$f$ line flux. 
  


In both these last two models, we have encountered a problem when we have adopted
a stellar inclination angle of $i \approx 21.5^{\circ}$ \citep{corder05} and
$v\sin i \approx 80$~km s$^{-1}$ \citep{boehm93}. The rotation period would then 
be 12.9 hr. This value is significantly  smaller than both the Mg\,{\sc ii} and
X-ray period ($\approx$ 42 hr) and the value suggested by \citet{catala99} for
the rotation period (32-34 hr). Further, 
in the hypothesis of a stellar corona, the latter should in this case not
exceed a height of 0.68~$R$  (the location of the co-rotation 
radius where centrifugal forces $\Omega d^2$ equal gravitational forces $GM/d^{2}$) 
above the surface at the equator, because otherwise centrifugal 
forces would make loops unstable \citep{collier88}. 
Further, with the adopted small inclination angle, no significant rotational  modulation 
should occur, and second, the $f/i$ ratio is at risk of being significantly suppressed.


The problems with a  wind in a magnetosphere or a corona would be alleviated if the 
stellar rotation axis were, for some reason, inclined against the axis of the circumstellar
disk, or the latter were largely warped between the inner regions and the well-observed outer
regions \citep{corder05}. The 33~hr period suggested by \citet{catala99} to be identified
with the rotation period would then require a stellar inclination of about 70~$\deg$, making partial
eclipses of X-ray emitting material, located for example close to the equatorial plane of the star,
easily possible. At the same time, azimuthally varying wind velocities would produce the
line shift periodicity  as reported in the optical and UV \citep{praderie86, catala86}. We have, however, 
no explanation  as to why the rotation axis should be inclined against the disk.

We further note that HD 163296 displays very similar properties to AB Aur (Table~\ref{tab2}), with a soft X-ray spectrum  
similar to the one that we have found for AB Aur.
However, the X-ray properties found for AB Aur and  HD 163296 are not common to all Herbig stars.
Some of these stars in fact display harder spectra,
with temperatures reaching several tens of MK and with larger X-ray luminosities
\citep{hamaguchi05,skinner04, stelzer06a}.
In fact, given the peculiar properties of the X-rays of AB Aur and the similar HD 163296 (Table~\ref{tab2},
\citealt{swartz05}), namely a very soft X-ray spectrum with a moderate $L_{\rm X}$, we suggest that
{\it these properties are defining properties of X-ray emission intrinsic to Herbig
stars}, while hard, flaring emission may be due to undetected companions (see
Sect.~\ref{introduction}, and Sect.~\ref{binaries}).

\begin{acknowledgements}
We acknowledge helpful comments by the referee.
We thank Laurence DeWarf, Edward Fitzpatrick, and Claude Catala for important information on fundamental
properties of AB Aur and Beate Stelzer for her helpful suggestions. This publication makes use of data products from the
Two Micron All Sky Survey, which is a joint project of the University of Massachusetts
and the Infrared Processing and Analysis Center/California Institute of Technology,
funded by the National Aeronautics and Space Administration and the National Science
Foundation.
Our research is based on observations obtained with {\it XMM-Newton}, an ESA science
mission with instruments and contributions directly funded by ESA Member States and 
the USA (NASA).
We would like to thank the International Space Science Institute (ISSI) in Bern,
Switzerland, for logistic and financial support during several workshops on the TMC XEST campaign.
X-ray astronomy research at PSI has been supported by the Swiss National Science
Foundation (grant 20-66875.01 and 20-109255/1). M.A. acknowledges support by NASA grant NNG05GF92G. 

\end{acknowledgements}

\end{document}